\documentclass[amsmath,amssymb,amsopn,superscriptaddress,preprint,author-numerical]{revtex4-1}

\usepackage{graphics,color}
\usepackage{textcomp}
\usepackage{braket,slashed}
\usepackage{hyperref}
\usepackage{enumerate}

\usepackage{bm}

\usepackage[mathscr]{euscript}
\usepackage{dsfont}

\newcommand{\operator}[1]{\ensuremath{\hat{#1}}}
\newcommand{\voperator}[1]{\ensuremath{\mathbb{#1}}}
\newcommand{\mset}[1]{\ensuremath{\mathcal{#1}}}
\newcommand{\group}[1]{\ensuremath{\mathsf{#1}}}
\newcommand{\vectorspace}[1]{\ensuremath{\mathbb{#1}}}
\newcommand{\algebra}[1]{\ensuremath{\mathfrak{#1}}}

\DeclareMathOperator{\order}{\mathscr{O}}
\DeclareMathOperator{\tr}{tr}

\DeclareMathOperator{\End}{\mathbb{L}}

\newcommand{\one}{\openone}
\renewcommand{\d}{\ensuremath{\mathrm{d}}}
\newcommand{\ic}{\ensuremath{\mathrm{i}}}
\newcommand{\ec}{\ensuremath{\mathrm{e}}}
\newcommand{\defis}{\ensuremath{\triangleq}}

\newcommand{\ham}{\ensuremath{\operator{H}}}
\newcommand{\hpsi}{\ensuremath{\hat{\psi}}}
\newcommand{\hpsid}{\ensuremath{\hat{\psi}^\dagger}}
\newcommand{\hPsi}{\ensuremath{\hat{\varPsi}}}
\newcommand{\hPsid}{\ensuremath{\hat{\varPsi}^\dagger}}

\newcommand{\rket}[1]{\ensuremath{|#1)}}
\newcommand{\rbra}[1]{\ensuremath{(#1|}}
\newcommand{\rbraket}[1]{\ensuremath{(#1)}}

\newcommand{\Rbraket}[1]{\ensuremath{\left(#1\right)}}
\newcommand{\Pexp}{\ensuremath{\mathscr{P}\exp}}

\newcommand{\hilbert}{\ensuremath{\vectorspace{H}}}
\newcommand{\Tplane}{\ensuremath{\vectorspace{T}}}
\newcommand{\varM}{\ensuremath{\mset{M}}}

\usepackage{amsthm}

\theoremstyle{remark}

\theoremstyle{definition}

\begin{document}

\title{Calculus of continuous matrix product states}
\author{Jutho Haegeman}
\email{jutho.haegeman@ugent.be}
\affiliation{Faculty of Physics and Astronomy, University of Ghent, Krijgslaan 281 S9, B-9000 Gent, Belgium}
\author{J.~Ignacio Cirac}
\affiliation{Max-Planck-Institut f\"ur Quantenoptik, Hans-Kopfermann-Str. 1, D-85748 Garching, Germany}
\author{Tobias J. Osborne}
\affiliation{Leibniz Universit\"{a}t Hannover, Institute of Theoretical Physics, Appelstrasse 2, D-30167 Hannover, Germany}
\affiliation{Leibniz Universit\"{a}t Hannover, Riemann Center for Geometry and Physics, Appelstrasse 2, D-30167 Hannover, Germany}
\author{Frank Verstraete}
\affiliation{Faculty of Physics and Astronomy, University of Ghent, Krijgslaan 281 S9, B-9000 Gent, Belgium}
\affiliation{Vienna Center for Quantum Science and Technology, Faculty of Physics, University of Vienna, Austria}

\begin{abstract}
We discuss various properties of the variational class of continuous matrix product states, a class of ansatz states for one-dimensional quantum fields that was recently introduced as the direct continuum limit of the highly successful class of matrix product states. We discuss both attributes of the physical states, \textit{e.g.} by showing in detail how to compute expectation values, as well as properties intrinsic to the representation itself, such as the gauge freedom. We consider general translation non-invariant systems made of several particle species and derive certain regularity properties that need to be satisfied by the variational parameters. We also devote a section to the translation invariant setting in the thermodynamic limit and show how continuous matrix product states possess an intrinsic ultraviolet cutoff. Finally, we introduce a new set of states which are tangent to the original set of continuous matrix product states. For the case of matrix product states, this construction has recently proven relevant in the development of new algorithms for studying time evolution and elementary excitations of quantum spin chains. We thus lay the foundation for similar developments for one-dimensional quantum fields. 
\end{abstract}

\maketitle

\tableofcontents
\clearpage

\section{Introduction}
Many revolutions and breakthroughs in quantum physics, and quantum many body physics in particular, were stimulated by guessing a suitable variational ansatz that captures the relevant correlations for the systems under consideration. Feynman's ansatz for the roton in superfluid Helium\cite{Feynman:1954aa,Feynman:1956aa}, the Bardeen-Cooper-Schrieffer wave function for superconductivity\cite{1957PhRv..106..162B} and the Laughlin wave function for the fractional quantum Hall effect\cite{PhysRevLett.50.1395} are only a few prominent examples. For gapped one-dimensional quantum spin systems, the set of matrix product states\cite{1987PhRvL..59..799A,1988CMaPh.115..477A,1992CMaPh.144..443F,2008AdPhy..57..143V,2009JPhA...42X4004C} is a very general ansatz that can describe a range of different phenomena and different physical phases, including normal symmetric and symmetry broken phases as well as the more exotic symmetry-protected topologically ordered phases such as the Haldane phase\cite{Haldane:1983aa,Haldane:1983ab,2010PhRvB..81f4439P}. Indeed, with the benefit of hindsight, we now understand White's powerful density matrix renormalization group algorithm\cite{1992PhRvL..69.2863W,1993PhRvB..4810345W} as a variational optimization over the set of matrix product states\cite{1995PhRvL..75.3537O,1997PhRvB..55.2164R}.

Until recently, few equally general ansatzes that surpass mean field theory were available for extended quantum systems in the continuum, \textit{i.e.} quantum fields. Numerical approaches require a finite number of degrees of freedom in order to fit the problem in the memory of a computer. For compact systems such as nuclei, atoms and molecules, an expansion in terms of a finite-dimensional basis is possible, but for extended systems this eventually results in a discretization to an effective lattice system. A new variational ansatz field theories in $d=1$ spatial dimensions was developed by Verstraete and Cirac in 2010 \cite{2010PhRvL.104s0405V}. This ansatz is formulated in the continuum and does not require an underlying lattice approximation. It can be considered to be the continuum limit of a special subclass of matrix product states (MPS) and is therefore called the \emph{continuous matrix product state} (cMPS) class.

The aim of the current paper is to discuss in greater detail the properties of cMPS. Section~\ref{s:def} reviews the different definitions and representations of these states in the current literature. We then derive a set of regularity conditions that become relevant in the case of systems with multiple particle species in Section~\ref{s:regularity}. Section~\ref{s:expectval} discusses how to (efficiently) evaluate expectation values with respect to these states. Section~\ref{s:gauge} is devoted to the gauge invariance and the existence of canonical forms in the continuous matrix product state representation for generic systems without translation invariance. We also discuss uniform continuous matrix product states in the thermodynamic limit and illustrate how continuous matrix product states possess a natural ultraviolet cutoff in Section~\ref{s:ti}. Finally, Section~\ref{s:tangent} provides an intuitive construction of tangent vectors to the variational set and discusses their representation properties as well, both for finite systems and in the thermodynamic limit. These tangent states are relevant when studying time evolution or elementary excitations along the lines of analogous MPS algorithms \cite{2011arXiv1103.0936H,2011arXiv1103.2286H,2012PhRvB..85c5130P,2012arXiv1207.0691M}. We do not strive for absolute mathematical rigor, but merely attempt to explain in full detail the prerequisites for using cMPS in numerical algorithms. For example, due to the intrinsic difficulty of the various infinite-dimensional function spaces involved, we do not include a rigorous proof that the set of continuous matrix product states constitutes a smooth (complex) manifold and that the construction of a tangent space is justified.

\section{Various definitions of the variational class}
\label{s:def}
\subsection{Setting}
\label{ss:def:setting}
Consider a quantum system defined on a one-dimensional continuum $\mset{R}=[-L/2,+L/2]$ with length $\lvert\mset{R}\rvert=L$ that accommodates $q$ bosonic and/or fermionic particle species, which are labeled by the greek index $\alpha=1,\ldots,q$. Throughout this paper, we restrict to non-relativistic systems. A state of the quantum system containing $N_{\alpha}$ particles of type $\alpha$ is then described by a square integrable function on $\prod_{\alpha=1}^{q}\mset{R}^{(N_{\alpha})}_{\eta_{\alpha}}$, where $\eta_{\alpha}=+1$  ($-1$) if particle species $\alpha$ is bosonic (fermionic) and $\mset{R}^{(N_{\alpha})}_{+}$ ($\mset{R}^{(N_{\alpha})}_{-}$) corresponds to the symmetric (antisymmetric) subspace of $\mset{R}^N$, the Cartesian product of $N$ copies of $\mset{R}$. The space of the square integrable functions on this domain is a Hilbert space that is denoted as
\begin{equation}
\hilbert_{\mset{R}}^{\{N_{\alpha}\}_{\alpha=1,\ldots,q}}=L^2\left(\prod_{\alpha=1}^{q}\mset{R}^{(N_{\alpha})}_{\eta_{\alpha}}\right).\label{eq:defNalphaspace}
\end{equation}
Following the principles of second quantization, we now define the Fock space
\begin{equation}
\hilbert_{\mset{R}}^{(\text{F})}=\bigoplus_{N_1=0}^{+\infty}\cdots \bigoplus_{N_q=0}^{+\infty}\hilbert_{\mset{R}}^{\{N_{\alpha}\}_{\alpha=1,\ldots,q}}\label{eq:deffockspace}
\end{equation}
which captures an arbitrary state of the quantum system. In addition, we denote the unique vacuum state as $\ket{\Omega}\in \hilbert_{\mset{R}}^{\{N_\alpha=0\}_{\alpha=1,\ldots,q}}$. Particles of type $\alpha$ are created and annihilated at position $x\in\mset{R}$ with the operators $\hpsid_{\alpha}(x)$ and $\hpsi_{\alpha}(x)$ with $\alpha=1,\ldots,q$. These satisfy the general commutation or anticommutation relations
\begin{align}
\hpsi_{\alpha}(x)\hpsi_{\beta}(y)-\eta_{\alpha,\beta} \hpsi_{\beta}(y)\hpsi_{\alpha}(x)&=0,&\hpsi_{\alpha}(x)\hpsid_{\beta}(y)-\eta_{\alpha,\beta} \hpsid_{\beta}(y)\hpsi_{\alpha}(x)&=\delta_{\alpha,\beta}\delta(x-y),\label{eq:commrelations}
\end{align}
where $\eta_{\alpha,\beta}=-1$ if both $\alpha$ and $\beta$ represent fermionic particles and $\eta_{\alpha,\beta}=1$ when at least one of the two particles species $\alpha$ or $\beta$ is bosonic. Clearly $\eta_{\alpha,\alpha}=\eta_{\alpha}$. We always write sums over the species index $\alpha$ explicitly and do not use Einstein's summation convention with respect to this index.

\subsection{Original definition}
\label{ss:def:original}
A cMPS is defined to be the state \cite{2010PhRvL.104s0405V}
\begin{multline}
\ket{\Psi[Q,R_{1},\ldots,R_{q}]}\defis
\tr\left(B \mathscr{P}\!\exp\left[\int_{-L/2}^{+L/2}\d x\, Q(x)\otimes \operator{\one}+\sum_{\alpha=1}^{q}R_{\alpha}(x) \otimes \hpsid_{\alpha}(x) \right]\right)\ket{\Omega},\label{eq:defcmps}
\end{multline}
where $\mathscr{P}\!\exp$ is the path ordered exponential (that orders its argument from left to right for increasing values of $x$) and $\ket{\Omega}$ is the empty vacuum that is annihilated by $\hpsi_{\alpha}(x)$, $\forall \alpha=1,\ldots,N$. The trace operation acts on an auxiliary space $\mathbb{C}^D$, also called the ancilla space, where $D$ is the bond dimension. The variational parameters correspond to the functions $Q, R_{\alpha}: \mset{R}\to \mathbb{C}^{D\times D}$ that take value in $\mathbb{L}(\mathbb{C}^D)\defis \mathbb{C}^{D\times D}$, the space of linear operators acting on the ancilla space. For now, we do not impose any continuity or regularity conditions on these functions, and we refer to Section~\ref{s:regularity} for a detailed discussion. Finally, the boundary operator $B\in \mathbb{L}(\mathbb{C}^D)$ encodes the boundary conditions. For a system with periodic boundary conditions the boundary operator has full rank and is typically chosen to be $B=\one_{D}$. In case of open boundary conditions, we can choose $B=\bm{v}_{\mathrm{R}}\bm{v}^{\dagger}_{\mathrm{L}}$ with $\bm{v}_{\mathrm{L}}$ and $\bm{v}_{\mathrm{R}}$ $D$-dimensional boundary vectors. Note that the matrix functions $Q$ and $R_{\alpha}$ themselves need to satisfy certain boundary conditions which are imposed by the physical setting. We discuss this in more detail in Section~\ref{s:bc}.

More formally, we can identify the cMPS construction as a map between the function spaces $\mset{R}\to \mathbb{C}^{D\times D}$ and the Fock space $\hilbert_{\mset{R}}^{(\text{F})}$:
\begin{equation}
\begin{split}
\Psi:&(\mset{R}\to \mathbb{C}^{D\times D})^{q+1} \to \hilbert_{\mset{R}}^{(\text{F})}:\\
&\qquad(Q,R_1,\ldots,R_q)\mapsto \ket{\Psi[Q,R_1,\ldots,R_q]}.
\end{split}
\end{equation}
The range of the map $\Psi$ defines a variational set $\mset{V}_{\mathrm{cMPS}(D)}\subset \hilbert_{\mset{R}}^{(\text{F})}$, where we often omit the explicit specification of the bond dimension. Henceforth, we compactly denote a cMPS $\ket{\Psi[Q,R_{1},\ldots,R_{q}]}$ as $\ket{\Psi[Q,\{R_{\alpha}\}]}$. It will always be clear from the context how many and which particle species are present. The variational set $\mset{V}_{\text{cMPS}(D)}$ is not a vector space, since the representation of the sum of two elements $\ket{\Psi[Q,\{R_{\alpha}\}]}+\ket{\Psi[Q',\{R_{\alpha}'\}]}$ requires in the most general case a cMPS $\ket{\tilde{\Psi}[\tilde{Q},\{\tilde{R}_{\alpha}\}]}\in\varM_{\text{cMPS}(\tilde{D})}$ with bond dimension $\tilde{D}=2D$, where we choose ($\forall x\in[-L/2,+L/2]$)
\begin{align*}
\tilde{Q}(x)&=Q(x)\oplus Q'(x),\\
\tilde{R}_{\alpha}(x)&=R_{\alpha}(x)\oplus R_{\alpha}'(x),&\forall \alpha=1,\ldots,q\\
\tilde{B}&=B\oplus B'.
\end{align*}
The variational set does however contain almost complete rays of states, since for any state $\ket{\Psi[Q,\{R_{\alpha}\}]}\in\mset{V}_{\text{cMPS}(D)}$ and any $\lambda\in\mathbb{C}_{0}=\mathbb{C}\setminus\{0\}$ we can also represent $\lambda\ket{\Psi[Q,\{R_{\alpha}\}]}$ as a cMPS with bond dimension $D$ as $\ket{\Psi[Q',\{R'_{\alpha}\}]}$, where $Q'(x)=Q(x)+\mu(x) \one_{D}$ and $R_{\alpha}'(x)=R_{\alpha}(x)$ with
\begin{displaymath}
\exp\left(\int_{-L/2}^{+L/2}\d x\,\mu(x)\right)=\lambda.
\end{displaymath}
A special case is obtained for $\lambda=0$, since this requires us to redefine $Q(x)$ as $Q'(x)=Q(x)-\infty \one_{D}$. Hence, the null state is not contained within $\mset{V}_{\text{cMPS}(D)}$ but only in its closure. Correspondingly, the variational set $\mset{V}_{\text{cMPS}(D')}$ with $D'<D$ is not a subset of $\mset{V}_{\text{cMPS}(D)}$. For example, if the boundary matrices are fixed to $B'=\one_{D'}$ and $B=\one_{D}$ (periodic boundary conditions), then a representation of the cMPS $\ket{\Psi'[Q',\{R_{\alpha}'\}]}$ with bond dimension $D'$ as a cMPS $\ket{\Psi[Q,\{R_{\alpha}\}]}$ with bond dimension $D>D'$ requires $Q=Q'\oplus (-\infty \times \one_{D-D'})$ and $R_{\alpha}=R_{\alpha}'\oplus (0\times \one_{D-D'})$, hence $\mset{V}_{\text{cMPS}(D')}$ is only included in the closure of $\mset{V}_{\text{cMPS}(D)}$. Note that this differs from the case of MPS on the lattice, where $\mset{V}_{\text{MPS}(D')}\subset \mset{V}_{\text{MPS}(D)}$ for $D\geq D'$. 

\subsection{Fock space embedding}
\label{ss:def:fockembedding}
The embedding of $\ket{\Psi[Q,\{R_{\alpha}\}]}\in\mset{V}_{\text{cMPS}(D)}$ in the Fock space $\hilbert_{\mset{R}}^{(\text{F})}$ for finite $\lvert\mset{R}\rvert$ can be made explicit by expanding the path ordered exponential as
\begin{multline*}
\ket{\Psi[Q,\{R_{\alpha}\}]}=\sum_{N=0}^{+\infty} \int_{-L/2\leq x_{1}\leq \cdots \leq x_{N}\leq L/2}\d x_{1}\cdots \d x_{N}\\
\tr\Bigg[ B \bigg(Q(x_1)\otimes \operator{\one}+\sum_{\alpha_1=1}^{q}R_{\alpha_1}(x_1) \otimes \hpsid_{\alpha_1}(x_1) \bigg)\times \cdots\\
\times \bigg(Q(x_N)\otimes \operator{\one}+\sum_{\alpha_N=1}^{q}R_{\alpha_N}(x_N) \otimes \hpsid_{\alpha_N}(x_N) \bigg)\Bigg]\ket{\Omega}.
\end{multline*}
We can then expand the round brackets and reorder the sum in terms of the actual number of created particles by grouping subsequent occurrences of the $Q$ term, so as to obtain
\begin{multline}
\ket{\Psi[Q,\{R_{\alpha}\}]}=\sum_{N=0}^{+\infty} \sum_{\alpha_1,\ldots,\alpha_N=1}^{q} \int_{-L/2\leq x_{1}\leq \cdots \leq x_{N}\leq L/2}\d x_{1}\cdots \d x_{N}\\
\tr\bigg[ B M_Q(-L/2,x_1) R_{\alpha_1}(x_1) M_Q(x_1,x_2) \cdots R_{\alpha_N}(x_N) M_Q(x_N,L/2) \bigg]\\
\hpsid_{\alpha_1}(x_1)\hpsid_{\alpha_2}(x_2)\cdots \hpsid_{\alpha_N}(x_N)\ket{\Omega},\label{eq:cmpsfockembedding}
\end{multline}
with
\begin{displaymath}
M_Q(x,y)=\sum_{k=0}^{+\infty} \int_{x\leq z_1\leq \cdots \leq z_k \leq y} \d z_1\cdots \d z_k Q(z_1) \cdots Q(z_k)= \mathscr{P}\ec^{\int_{x}^{y} Q(z) \d z}.
\end{displaymath}
Eq.~\eqref{eq:cmpsfockembedding} shows how a cMPS can be interpreted as an superposition over the different particle number sectors in the Fock space. Note that this is not completely equivalent to the different sectors $\hilbert_{\mset{R}}^{\{N_{\alpha}\}_{\alpha=1,\ldots,q}}$ in the direct product construction of $\hilbert_{\mset{R}}^{(\text{F})}$ [Eq.~\eqref{eq:deffockspace}], since now only the total number of particles $N=\sum_{\alpha=1}^{q} N_{\alpha}$ is fixed. If we define the $N$-particle wave functions as
\begin{equation}
\phi_{\alpha_{1},\ldots,\alpha_{N}}(x_{1},\ldots,x_{N})=\braket{\Omega|\hpsi_{\alpha_{k}}(x_{k})\cdots \hpsi_{\alpha_{1}}(x_{1})|\Psi[Q,\{R_{\alpha}\}]}.\label{eq:defphiN}
\end{equation}
then we can infer from Eq.~\eqref{eq:cmpsfockembedding} that
\begin{multline}
\phi_{\alpha_{1},\ldots,\alpha_{N}}(x_{1},\ldots,x_{N}) =\\
\tr\bigg[ B M_Q(-L/2,x_1) R_{\alpha_1}(x_1) M_Q(x_1,x_2) \cdots R_{\alpha_N}(x_N) M_Q(x_N,L/2) \bigg]\label{eq:cmpsNparticle}
\end{multline}
only when $x_1\leq x_2\leq \cdots \leq x_N$. It can be extended to any other order of the arguments by reordering the annihilation operators in Eq.~\eqref{eq:defphiN} according to the given commutation or anticommutation relations in Eq.~\eqref{eq:commrelations}. The non-relativistic kinetic energy requires that these functions are sufficiently regular, which together with the extension to arbitrary order of the arguments imposes certain non-trivial constraints on the matrix functions $Q$ and $R_{\alpha}$ that are to be discussed in Section~\ref{s:regularity}.

\subsection{The continuum limit of matrix product states}
\label{ss:def:continuum}
The cMPS $\ket{\Psi[Q,\{R_{\alpha}\}]}$ was originally constructed in Ref.~\onlinecite{2010PhRvL.104s0405V} as the continuum limit of a certain subset of MPS, where the subset was selected in such a way as to obtain a valid continuum limit. We explore this construction in greater detail and elaborate on some of the non-trivial implications regarding ultraviolet cutoffs and correlation lengths (infrared cutoffs).

We approximate the continuum $\mset{R}=[-L/2,L/2]$ by a lattice $\mset{L}$ with lattice spacing $a$ and $N=L/a$ sites, where we send $a\to 0$. On every site of the lattice we can create and annihilate particles of type $\alpha$ by acting with the creation and annihilation operators $\operator{c}_{\alpha}^{\dagger}(n)$ and $\operator{c}_{\alpha}(n)$. We can relate them to the field operators by
\begin{align}
\operator{c}_{\alpha}(n)=\int_{na}^{(n+1) a} \hpsi_{\alpha}(x)\, \d x
\end{align}
and its hermitian conjugate. The local basis on site $n$ thus consists of the states $\ket{0}_{n}$ (no particles), $\ket{\alpha}_{n}=c_{\alpha}^{\dagger}(n)\ket{0}_{n}$, $\ket{\alpha,\beta}_{n}=c_{\alpha}^{\dagger}(n)c_{\beta}^{\dagger}(n)\ket{0}_{n}$, \ldots\ On this lattice, we can define an MPS $\ket{\Psi[A]}$ with matrices $A^{s}(n)$ where $s$ can take values $0$, $\alpha$, $(\alpha,\beta)$, \ldots\ If the local basis is infinite-dimensional, this MPS definition is only formal, \textit{i.e.}\ it cannot be used for practical computations. In the limit $a\to 0$, the number of sites $L/a$ in the lattice $\mset{L}$ goes to infinity.

On an infinite number of lattice sites, two arbitrary MPS are generally orthogonal due to the (infrared) orthogonality catastrophe\cite{Anderson:1967aa}. Since we now aim to create quantum field states within the Fock space $\hilbert_{\mset{R}}^{(\text{F})}$, we need to restrict to a special subset of MPS where the total number of particles is finite (on average, so that $\braket{ \operator{N}}$ is finite). Since a finite number of particles has to be distributed over a diverging number of sites $L/a$, most of the sites in the lattice $\mset{L}$ are empty on average. So $A^{0}$ has to be the dominant matrix, and it turns out that the cMPS $\ket{\Psi[Q,\{R_{\alpha}\}]}\in\hilbert_{\mset{R}}^{(\text{F})}$ can be obtained from the continuum limit ($a\to 0$) of the MPS $\ket{\Psi[A]}\in\hilbert_{\mset{L}}$ by identifying $\hpsid_{\alpha}(n a)=\operator{c}^{\dagger}_{\alpha}(n)/\sqrt{a}$ and
\begin{align}
A^{0}(n)&=\one_{D}+a Q(n a),\nonumber\\
A^{\alpha}(n) &= \sqrt{a} R_{\alpha}(n a),\nonumber\\
A^{(\alpha,\beta)}(n) &= \begin{cases} \frac{a}{2} [ R_{\alpha}(n a) R_{\beta}(n a)+\eta_{\alpha,\beta} R_{\beta}(n a) R_{\alpha}(n a)],& \alpha\neq \beta\\
\frac{a}{2} R_{\alpha}(n a)^{2},&\alpha=\beta
\end{cases}\label{eq:correspondencemps}\\
&\ldots\nonumber
\end{align}
together with $\ket{\Omega}=\ket{\bm{0}}=\otimes_{n\in\mset{L}} \ket{0}_{n}$, $\forall n=-L/2a,-L/2a+1,\ldots,+L/2a-1$. This equivalence can be obtained from a Taylor expansion of the $\exp$-operator, although this is only completely rigorous when the entries of $Q$ and $R_{\alpha}$ are finite and the operators $\hpsid(x)$ are bounded (\textit{i.e.} not for bosons). Most results for cMPS in the remainder of this chapter can be derived from this correspondence with MPS, but we attempt to derive these results directly in the continuum as much as possible.

The correspondence with MPS is useful for concluding that the entanglement of one half of the chain with the other half (in the case of open boundary conditions) is limited by the upper bound $\log D$. By restricting to MPS within a single Fock space in the thermodynamic limit, we avoid the orthogonality catastrophe. The infrared orthogonality catastrophe of MPS in the thermodynamic limit would turn into an ultraviolet catastrophe when this infinitely-sized lattice $\mset{L}$ would correspond to the continuum limit of a finitely sized continuum $\mset{R}$. Physically, the ultraviolet catastrophe is avoided because the finite number of particles induce a physical cutoff $a_{\text{phys}}$ that is given, not by the lattice spacing $a\to 0$ but by $a_{\text{phys}}=\rho^{-1}$ with $\rho=\braket{\operator{N}}/L$ the particle density\footnote{cMPS still obey the infrared orthogonality catastrophe when formulated in the thermodynamic limit (see Section~\ref{s:ti})}. The presence of a physical length scale can be detected from the physical dimensions of $Q$ and $R_{\alpha}$, which are given by $[Q]=\ell^{-1}$ and $[R]=\ell^{-1/2}$ with $\ell$ a generic length dimension. The nature of the physical cutoff $a_{\text{phys}}$ and its relation to $Q$ and $R_{\alpha}$ is discussed in Section~\ref{s:ti} for the translation invariant case, where it can unambiguously be defined. Shifting the cutoff from the lattice spacing $a$ to a physical value $a_{\text{phys}}$ is a very important step in the definition of cMPS. MPS with finite bond dimension $D$ have a finite amount of entanglement to which corresponds in general a finite range of fluctuations $\xi/a$, where $\xi$ denotes the correlation length. Hence, they have in general a finite dimensionless correlation length $\tilde{\xi}=\xi/a$. As $a$ is scaled to zero while $\tilde{\xi}$ remains finite, the physical correlation length $\xi$ would also scale to zero. It is because the physical cutoff is shifted to a finite value $a_{\text{phys}}$ (with thus $a_{\text{phys}}/a\to \infty$) that cMPS are able to combine a finite amount of entanglement with a finite physical correlation length $\xi$ (with thus $\xi/a\to \infty$ but with $\xi/a_{\text{phys}}$ finite). The physical correlation length $\xi$ is also computed in Section~\ref{s:ti} for the translation invariant case. 

\subsection{Alternative construction through continuous measurement}
\label{ss:def:continuousmeasurement}
Rather than trying to construct a cMPS as the continuum limit of a MPS, we could also try to directly define the continuum limit of the processes that define MPS. Unfortunately, the process of sequential Schmidt decompositions has no straightforward generalization to the continuum and neither has the definition of valence bond solids. One can however define a continuum version of the sequential generation process that creates MPS\cite{2005PhRvL..95k0503S}, based on the paradigm of continuous measurement \cite{Caves:1987aa}. The resulting process for creating cMPS is described in Ref.~\onlinecite{2010PhRvL.105z0401O}, and is here summarised for the sake of completeness.

As in the discrete case, let the ancilla start in a state $\bm{v}_{\text{R}}\in\hilbert_{\text{ancilla}}=\mathbb{C}^{D}$. This ancilla can be interpreted as a resonating cavity with $D$ internal levels, in which there is a particle source that creates particles of type $\alpha$ ($\alpha=1,\ldots,q$). These particles gradually leave the cavity due to cavity losses. Since particles leaving the cavity at different times occupy different positions in space at a given time (since they travel at a certain speed which we set equal to one), the resulting configuration of particles can be interpreted as a static spatially distributed quantum state. For a compact cavity (\textit{i.e.} a zero-dimensional system), the resulting quantum state is one-dimensional. As an abstraction of this physical process, a $(d-1)$-dimensional cavity can be used to encode a $d$-dimensional holographic quantum state. We refer to Ref.~\onlinecite{2010PhRvL.105z0401O} for the general case, and henceforth restrict to the $d=1$ case that produces cMPS.

Between two particle emissions, the cavity evolves according to a Hamiltonian $K\in\End(\mathbb{C}^D)$ (a Hermitean $D\times D$ matrix), whereas the physical state outside the cavity does not evolve. By observing the particles that are emitted from the cavity, we are continuously measuring the state of the cavity (\textit{i.e.} ancilla). The state of the cavity at time $t$ is encoded in the particle distribution at position $x=-t$. It was shown that the resulting configuration of particles outside the cavity is given by
\begin{equation}
\bm{v}_{\mathrm{L}}^{\dagger}\Pexp\left(-\ic \int_{-L/2}^{+L/2}\d x\, K(x)\otimes\operator{\one} + \sum_{\alpha=1}^{N}\ic R_{\alpha}(x)\otimes \hpsid_{\alpha}(x)-\ic R_{\alpha}(x)^{\dagger}\otimes \hpsi_{\alpha}(x)\right) \bm{v}_{\mathrm{R}} \ket{\Omega},\label{eq:defcontmeasurement}
\end{equation}
where the ancilla is projected onto the state $\bm{v}_{\mathrm{L}}$ at the end of the measurement, in order to decouple it from the physical state. The resulting expression does not yet correspond exactly to Eq.~\eqref{eq:defcmps} but it can easily be brought in the required form by using the Baker-Campbell-Hausdorff formula on every infinitesimal patch of the path ordered exponential. We then obtain that the state in Eq.~\eqref{eq:defcontmeasurement} is contained within $\mset{V}_{\mathrm{cMPS}}$, as it is equal to $\ket{\Psi[Q,\{R_{\alpha}\}]}$ for the specific choice
\begin{align}
Q(x)=-\ic K(x) -\frac{1}{2}\sum_{\alpha=1}^{N} R_{\alpha}(x)^{\dagger}R_{\alpha}(x).\label{eq:qunitary}
\end{align}
We recall that $K(x)$ is a Hermitian matrix. Generic cMPS can be brought into this form by using the gauge invariance of the cMPS representation, as discussed in Section~\ref{s:gauge}.

This construction allows us to introduce a unitary operator $\operator{U}(y,z)\in\End(\mathbb{C}^{D}\otimes \hilbert)$
\begin{equation}
\operator{U}(y,z)=\Pexp\left(-\ic \int_{z}^{y}\d x\, K(x)\otimes\operator{\one} + \sum_{\alpha=1}^{N}\ic R_{\alpha}(x)\otimes \hpsid_{\alpha}(x)-\ic R_{\alpha}(x)^{\dagger}\otimes \hpsi_{\alpha}(x)\right).\label{eq:defUalternative}
\end{equation}
Being a unitary operator, it conserves the norm of $\bm{v}_{\mathrm{R}}\otimes\ket{\Omega}$. This does not imply that the cMPS $\ket{\Psi[Q,\{R_{\alpha}\}]}$ with $Q$ given by Eq.~\eqref{eq:qunitary} is automatically normalized to unity, because the definition also involves a projection to $\bm{v}_{\mathrm{L}}$. But the unitarity of $\operator{U}(y,z)$ in Eq.~\eqref{eq:defUalternative} does guarantee that $\ket{\Psi[Q,\{R_{\alpha}\}]}$ can easily be normalized and has no norm that diverges or goes to zero in the large volume limit.

From a physical perspective, this construction is important as it clearly sketches the holographic properties of the cMPS. The physical state of a one-dimensional system is described by a zero-dimensional boundary theory. The spatial coordinate of the physical system acts as a time coordinate in the boundary theory. The physical state is created because the boundary theory interacts with the physical system, where the position of the interaction shifts linearly in time. This interaction results in the boundary theory not being at equilibrium. Instead, the boundary theory is subject to dissipative dynamics, as will become clear in the following section. This holographic property is of course strongly related with the intrinsic area law for entanglement that is present in cMPS.

\subsection{Path integral representation}
\label{ss:def:pathintegral}
Recently, it has also been illustrated that we can break up the path ordered exponential in the definition of $\ket{\Psi[Q,\{R_\alpha\}]}$ and insert resolutions of the identity in order to obtain a path integral description of the same state\cite{Brockt:fk}. The easiest way to insert an identity is by first introducing a second quantized version of the ancilla by making the substitution
\begin{align}
Q(x)& \mapsto \hat{Q}(x)=Q^{j,k}(x) \hat{b}_j^\dagger \hat{b}_k,&R_{\alpha}(x) &\mapsto \hat{R}_{\alpha}(x)=R_{\alpha}^{j,k}(x) \hat{b}_j^\dagger \hat{b}_k,
\end{align}
with $\hat{b}_j$ and $\hat{b}^\dagger_j$ annihilation and creation operators for bosonic or fermionic particles in level $j=1,\ldots,D$ of the ancilla. The resolution of the identity can now be expressed in terms of coherent states. However, the ancilla Hilbert space is now an infinite-dimensional Fock space, whereas the original ancilla space was only $\mathbb{C}^D$ and corresponds to the single-particle sector of this Fock space. Because the operators $\hat{Q}(x)$ and $\hat{R}_{\alpha}(x)$ are particle-number preserving with respect to the ancilla, we can restrict the whole path integral to the single particle sector by either choosing appropriate boundary conditions. If $\ket{\omega}$ denotes the ancilla zero-particle state, then a restriction to the single particle state is obtained by identifying
\begin{align}
B&\mapsto \hat{B}=B^{j,k} b^\dagger_j \ket{\omega}\bra{\omega} b_k.
\end{align}
If we introduce the coherent states
\begin{equation}
\ket{\phi}=\exp\left(\sum_{j=1}^{D} \phi_j \hat{b}^{\dagger}_j - \phi^\ast_j \hat{b}_j\right)\ket{\omega}
\end{equation}
then we can write the identity as
\begin{equation}
\hat{\one}=\frac{1}{\pi^D} \int \prod_{j=1}^D \d \phi_j\d \phi_j^\ast \, \ket{\phi}\bra{\phi}.
\end{equation}
Following the standard recipe, we can then obtain the path integral description of $\ket{\Psi[Q,\{R_{\alpha}\}]}$ as
\begin{multline}
\ket{\Psi[Q,\{R_{\alpha}\}]}=\\
\int \mathscr{D} \phi \mathscr{D}\phi^{\ast} \left(\phi(+L/2)^\dagger B \phi(-L/2)\right) \ec^{-\frac{\lvert \phi(-L/2)\rvert^2}{2}-\frac{\lvert \phi(L/2)\rvert^2}{2}}\qquad\qquad\qquad\qquad\qquad\qquad\\
\times \exp\bigg[\int_{-L/2}^{+L/2} \Big\{\frac{1}{2}\phi^\dagger(x)\frac{\d \phi}{\d x}(x) -\frac{1}{2} \frac{\d \phi^{\dagger}}{\d x}(x) \phi(x) + \phi^\dagger(x)Q(x)\phi(x)\\
+ \sum_{\alpha=1}^{q} \left(\phi^\dagger(x) R_{\alpha}(x)\phi(x)\right) \hpsid_\alpha(x)\Big\}\, \d x \bigg]\ket{\Omega},\label{eq:pathintegralrepresentation}
\end{multline}
where $\phi(x)$ is a $D$-dimensional vector function with components $\phi_j(x)$, $j=1,\ldots,D$. This path integral representation can serve as a useful starting point for generalizations of the cMPS, \textit{e.g.} by replacing the second quantized auxiliary system by a true field theory, so that this becomes the cMPS analogon of the construction in Ref.~\onlinecite{2010PhRvB..81j4431C,2011PhRvA..83e3807N}. If this field theory is a conformal field theory, it is then very close in spirit to some model states for Quantum Hall Systems\cite{Moore1991362,Dubail:fk}.

\section{Regularity conditions}
\label{s:regularity}
In Eq.~\eqref{eq:defphiN} we have defined the $N$-particle wave functions $\phi_{\alpha_{1},\ldots,\alpha_{N}}(x_{1},\ldots,x_{N})$. For $x_{1}\leq \cdots \leq x_{N}$ these are completely specified by Eq.~\eqref{eq:cmpsNparticle}. However, for general choices of the matrix functions $Q$ and $R_{\alpha}$, the extension of Eq.~\eqref{eq:cmpsNparticle} to all orders of its arguments does not automatically satisfy the required properties that a physical $N$-particle wave function should satisfy. For example, the $N$-particle wave functions should be differentiable in each of its arguments if the state has to produce a finite non-relativistic kinetic energy.

However, there is no need to work with the Fock space expansion of Eq.~\eqref{eq:cmpsfockembedding}. We can check the regularity of the $N$-particle wave functions by immediately evaluating the kinetic energy in second quantization. For further reference, we first define
\begin{align}
\operator{U}(x,y)=\mathscr{P} \exp\left[\int_{x}^{y}\d z\, \left\{Q(z)\otimes \operator{\one} + \sum_{\alpha=1}^{q}R_{\alpha}(z)\otimes \hpsid_{\alpha}(z)\right\}\right]\label{eq:defU},
\end{align}
where $\operator{U}(x,y)\in\End(\hilbert \otimes \mathbb{C}^{D})$ with $\mathbb{C}^{D}$ the ancilla space, \textit{i.e.} it is a $D\times D$ matrix of operators. Unlike the operator $\operator{U}(y,z)$ defined in Subsection~\ref{ss:def:fockembedding}, the operator in Eq.~\eqref{eq:defU} is not unitary. It only equals the unitary version when acting on $\ket{\Omega}$ and if $Q(z)$ is given by Eq.~\eqref{eq:qunitary}. In addition, we define a closely related set of operators $\operator{U}_\alpha(x,y)$ ($\alpha=1,\ldots,q$) as
\begin{equation}
\operator{U}_{\alpha}(x,y)=\mathscr{P} \exp\left[\int_{x}^{y}\d z\,\left\{ Q(z)\otimes \operator{\one} + \sum_{\beta=1}^{q}\eta_{\alpha,\beta}R_{\beta}(z)\otimes \hpsid_{\beta}(z)\right\}\right]\label{eq:defUalpha}.
\end{equation}
In order to compute any expectation value, which is the topic of the next section, we need to be able to act with the field annihilation operators $\hpsi_{\alpha}(x)$ on the state $\ket{\Psi[Q,\{R_{\alpha}\}]}$. If we are able to drag $\hpsi_{\alpha}(x)$ through the path-ordered exponential, it then acts on $\ket{\Omega}$, which is annihilated by any field operator. We can now use Eq.~\eqref{eq:commutatorequalitygeneralized} as derived in Appendix~\ref{a:formula}, where $\operator{B}=\hpsi_{\alpha}(x)$, $\operator{A}_1(z)$ contains both $Q(z)\otimes \operator{\one}$ and any term $R_{\beta}(z)\otimes \hpsid_{\beta}(z)$ for which $\eta_{\alpha,\beta}=1$, and $\operator{A}_2(z)$ contains the terms $R_{\beta}(z)\otimes \hpsid_{\beta}(z)$ for which $\eta_{\alpha,\beta}=-1$. We then obtain
\begin{displaymath}
\hpsi_{\alpha}(x)\operator{U}(-L/2,+L/2)-\operator{U}_{\alpha}(-L/2,+L/2)\hpsi_{\alpha}(x)=\operator{U}_{\alpha}(-L/2,x) R_{\alpha} \operator{U}(x,+L/2)
\end{displaymath}
which immediately results in
\begin{equation}
\hpsi_{\alpha}(x) \ket{\Psi[Q,\{R_{\beta}\}]}=\tr\left[B \operator{U}_{\alpha}(-L/2,x) R_{\alpha}(x) \operator{U}(x,+L/2)\right]\ket{\Omega}.\label{eq:psiPsi}
\end{equation}
Hence, acting with an annihilation operator of type $\alpha$ at position $x$ not only lowers a matrix $R_{\alpha}(x)$, but also transforms the path ordered exponential $\operator{U}(-L/2,x)$ into $\operator{U}_{\alpha}(-L/2,x)$, because we had to take the particle statistics into account for bringing $\hpsi_{\alpha}(x)$ to the position where it could lower $R_{\alpha}(x)$. 

The non-relativistic kinetic energy operator $\operator{T}$ is given by
\begin{equation}
\operator{T}=\int_{-L/2}^{+L/2}\operator{t}(x)\,\d x,
\end{equation}
where the kinetic energy density $\operator{t}(x)$ at position $x$ is given by
\begin{equation}
\operator{t}(x)=\sum_{\alpha=1}^{N} \frac{1}{2 m_{\alpha}} \left(\frac{\d \hpsid_{\alpha}}{\d x}(x)\right) \left(\frac{\d \hpsi_{\alpha}}{\d x}(x)\right).
\end{equation}
Hence, a finite kinetic energy expectation value $\braket{\Psi[\overline{Q},\{\overline{R}_{\alpha}\}]|\operator{T}|\Psi[Q,\{R_{\alpha}\}]}$ requires that the state $\frac{\d \hpsi_{\alpha}}{\d x}(x)\ket{\Psi[Q,\{R_{\alpha}\}]}$ has a finite norm. Differentiating Eq.~\eqref{eq:psiPsi} and using Eq.~\eqref{eq:diffU}, we obtain
\begin{align}
\frac{\d\ }{\d x}\hpsi_{\alpha}(x)& \ket{\Psi[Q,\{R_{\beta}\}]}\nonumber\\
=&\tr\Bigg[B \operator{V}_{\alpha}(-L/2,x) \bigg([Q(x),R_{\alpha}(x)]+\frac{\d R_{\alpha}}{\d x}(x)\bigg)\operator{U}(x,+L/2)\Bigg]\ket{\Omega}\nonumber\\
&+\tr\Bigg[B \operator{V}_{\alpha}(-L/2,x) \bigg(\sum_{\beta=1}^{q}\big[\eta_{\alpha,\beta} R_{\beta}(x)R_{\alpha}(x)\nonumber\\
&\qquad\qquad\qquad\qquad\qquad- R_{\alpha}(x)R_{\beta}(x)\big]\otimes\hpsid_{\beta}(x)\bigg)\operator{U}(x,+L/2)\Bigg]\ket{\Omega}.\label{eq:diffpsiPsi}
\end{align}
The term on the first line can be shown to have a finite norm (see next section), provided of course that $R_\alpha(x)$ is a differentiable function with a well-behaved derivative $\d R_\alpha(x)/d x$ at any $x\in\mset{R}$. Since the term on the second line of Eq.~\eqref{eq:diffpsiPsi} has particles of any species $\beta=1,\ldots,q$ being created at the fixed position $x$, this term is not normalizable. Put differently, $\lVert (\d \hpsi(x)/\d x)\ket{\Psi[Q,\{R_{\alpha}\}]}\rVert^{2}$ contains a divergent contribution $\delta(0)$ (in position space), unless we impose the \emph{regularity condition}
\begin{align}
\eta_{\alpha,\beta} R_{\beta}(x)R_{\alpha}(x) -R_{\alpha}(x) R_{\beta}(x)=0, \quad \forall x\in \mset{R}.\label{eq:regcondition}
\end{align}
Hence the matrices $R_{\alpha}$ should have the same statistics as the particle creation operators to which they couple. For systems with a single species of bosons, the condition in Eq.~\eqref{eq:regcondition} is automatically fulfilled. For systems with multiple species of bosons, it requires that any two matrices $R_{\alpha}(x)$ and $R_{\beta}(x)$ at the same spatial point $x$ commute. If $\alpha$ is a fermionic particle species, the corresponding matrix $R_{\alpha}(x)$ has to satisfy $R_{\alpha}(x)^{2}=0$, $\forall x\in\mset{R}$. When two particles of fermionic type $\alpha$ approach each other, there is a corresponding factor $R_{\alpha}(y) \Pexp(\int_{y}^{z}\d x\, Q(x)) R_{\alpha}(z)$ in the $N$-particle wave function $\phi_{\alpha_{1},\ldots,\alpha,\alpha,\ldots \alpha_{N}}(x_{1},\ldots,y,z,\ldots,x_{N})$. For $y\to z$, the exponential factor continuously evolves towards $\one_{D}$, so that the $k$-particle wave function continuously becomes zero. Hence, the finiteness of the kinetic energy requires that two fermionic particles of the same type cannot come arbitrarily close together and thus imposes Pauli's principle. 

Differentiability of the wave function is sufficient for a finite kinetic energy, which is by far the most important physical requirement of the wave function. We can also impose higher regularity constraints on the $N$-particle wave functions. Since these do in general not arise from physical considerations, we postpone this discussion to Appendix~\ref{a:higherorderregularity}. While the resulting conditions are interesting from an algebraic point of view, they are in general hard to satisfy with finite-dimensional matrices. For practical applications, satisfying the original condition in Eq.~\eqref{eq:regcondition}, as imposed by the finiteness of the kinetic energy, should be sufficient.

We conclude this subsection by investigating what else can be learned from the physical considerations concerning particle statistics. The regularity conditions [Eq.~\eqref{eq:regcondition}] already require that the matrices $R_{\alpha}$ behave as the corresponding operators $\hpsi_{\alpha}$ in terms of commutation and anticommutation relations. In a physical system, we should not have fermionic condensates, \textit{i.e.} $\braket{\Psi|\hpsi_{\alpha}(x)|\Psi}=0$ if particle species $\alpha$ is fermionic. This is a consequence of the invariance of an physical Hamiltonian $\ham$ under the action of the parity operator $\operator{P}$, which flips the sign of any fermionic operator ($\operator{P}\hpsi_{\alpha}(x)\operator{P}=\eta_{\alpha,\alpha}\hpsi_{\alpha}(x)$) and is thus idempotent ($\operator{P}=\operator{P}^{-1}=\operator{P}^{\dagger}$). We can construct $\operator{P}$ as
\begin{equation}
\operator{P}=\exp\left[\ic \pi \sum_{\alpha\ \text{fermionic}} \operator{N}_{\alpha}\right]=\exp\left[\ic \pi \sum_{\alpha\ \text{fermionic}} \int_{\mset{R}} \d x\, \hpsid_{\alpha}(x)\hpsi_{\alpha}(x)\right].
\end{equation}
Physical states satisfy $\operator{P}\ket{\Psi}=\ec^{\ic \phi} \ket{\Psi}$, where the idempotence of $\operator{P}$ requires $\phi=0$ or $\phi=\pi$. Physical states thus consist completely of a superposition of states, all of which have either an even or an odd number of fermions. Imposing this same property for cMPS requires one to explicitly incorporate the $\mathbb{Z}_{2}$ symmetry (with group elements $\{\operator{\one},\operator{P}\}$) in the matrix structure of $R_{\alpha}$ and $Q$. Since $\operator{P}\ket{\Psi[Q,\{R_{\alpha}\}]}=\ket{\Psi[Q,\{\eta_{\alpha,\alpha} R_{\alpha}\}]}$, we should also be able to define a virtual operator $P\in\End(\mathbb{C}^{D})$ such that $P Q P^{-1}=Q$ and $P R_{\alpha} P^{-1} =\eta_{\alpha,\alpha} R_{\alpha}$. This operator can in principle be $x$-dependent, but we should then be able to apply a local gauge transformation (see Section~\ref{s:gauge}) in order to make $P$ space-independent. In addition, it is clear from the definition that $P$ is idempotent ($P=P^{-1}$). If we can assume that $P$ is diagonalizable, then $P$ divides the ancilla space $\mathbb{C}^{D}$ into a sector with positive parity (eigenspace of eigenvalue $+1$) and a sector with negative parity (eigenspace of $-1$). A global gauge transformation brings $P$ into the diagonal form
\begin{equation}
P=\begin{bmatrix} \one_{D^{(+)}} & 0_{D^{(+)}\times D^{(-)}} \\ 0_{D^{(-)}\times D^{(+)}} & -\one_{D^{(-)}}\end{bmatrix}
\end{equation}
with $D^{(+)}+D^{(-)}=D$. The required transformation behavior of $Q$ and $R_{\alpha}$ then imposes the following decomposition
\begin{align}
Q&=\begin{bmatrix} Q^{(+)} & 0_{D^{(+)}\times D^{(-)}} \\ 0_{D^{(-)}\times D^{(+)}} & Q^{(-)}\end{bmatrix},\\
R_{\alpha}&=\begin{bmatrix} R_{\alpha}^{(+)} & 0_{D^{(+)}\times D^{(-)}} \\ 0_{D^{(-)}\times D^{(-)}} & R_{\alpha}^{(-)} \end{bmatrix}\qquad \text{(particle species $\alpha$ is bosonic)},\\
R_{\alpha}&=\begin{bmatrix} 0_{D^{(+)}\times D^{(+)}} & R_{\alpha}^{(+-)} \\ R_{\alpha}^{(-+)} & 0_{D^{(-)}\times D^{(-)}}\end{bmatrix}\qquad \text{(particle species $\alpha$ is fermionic)}.
\end{align}
In the cMPS $\ket{\Psi[Q,\{R_{\alpha}\}]}$, all contributions with either an even or an odd number of fermions in Eq.~\eqref{eq:cmpsfockembedding} drop out, depending on the boundary matrices $B$. If only states with an even number of fermions are allowed, $B$ should have a decomposition as
\begin{align}
B&=\begin{bmatrix} B^{(+)} & 0_{D^{(+)}\times D^{(-)}} \\ 0_{D^{(-)}\times D^{(+)}} & B^{(-)}\end{bmatrix},
\end{align}
whereas a decomposition of the form
\begin{align}
B&=\begin{bmatrix} 0_{D^{(+)}\times D^{(+)}} & B_{\alpha}^{(+-)} \\ B_{\alpha}^{(-+)} & 0_{D^{(-)}\times D^{(-)}}\end{bmatrix}\end{align}
is required to select only states with an odd number of fermions.

\section{Boundary conditions}
\label{s:bc}
We have already mentioned in Section~\ref{s:def} that the type of boundary conditions ---open or periodic--- is encoded in the rank of the boundary matrix $B$. For a system with periodic boundary conditions, $B$ has full rank and is typically chosen to be the identity ($B=\one_{D}$). Since periodic boundary conditions identify the points $x=-L/2$ and $x=+L/2$, it is natural to assume that the matrix functions $Q$ and $R_{\alpha}$ are also single-valued, \textit{i.e.} $Q(-L/2)=Q(+L/2)$ and $R_{\alpha}(-L/2)=R_{\alpha}(+L/2)$ for all $\alpha=1,\ldots,q$.

For a system with open boundary conditions, it is suitable to work with a boundary matrix of the form $B=\bm{v}_{\mathrm{R}}\bm{v}_{\mathrm{L}}^{\dagger}$, \textit{i.e.} the rank of $B$ is one. However, in the case of open boundary conditions physical requirements impose additional conditions on the $N$-particle wave functions of Eq.~\eqref{eq:cmpsNparticle}. Typically, a finite system is interpreted as being embedded in an infinite system and having an infinitely strong potential energy outside of the interval $\mset{R}$, \textit{i.e.} $v(x)=+\infty$ for $x<-L/2$ and $x>+L/2$. The single particle wave functions that build up the Fock space are zero outside $\mset{R}$. A finite kinetic energy imposes continuity, and thus requires that the single particle wave functions are zero at $x=\pm L/2$. Consequently, the resulting $N$-particle wave functions have to produce zero as soon as one of the arguments $x_i$ is equal to $\pm L/2$. Since this has to be true for any configuration of the remaining $N-1$ particles, we obtain that we have to impose
\begin{align}
\bm{v}_{\mathrm{L}}^\dagger R(-L/2)&=0 &\text{and}&&R(+L/2)\bm{v}_{\mathrm{R}}=0.\label{eq:qropenbc}
\end{align}
A more detailed discussion of these conditions is presented in Ref.~\cite{qgp}, where a partial differential equation for the evolution of $Q$ and $R_{\alpha}$ under real or imaginary time dynamics is derived. In order to solve this partial differential equation, it needs to be complemented by the proper boundary conditions as given above. Throughout the remainder of this manuscript, we assume that we are working with cMPS where the matrix functions $Q$ and $R_{\alpha}$ satisfy the required conditions.

We now also have to discuss whether we can completely fix the boundary matrix $B$, or whether its entries should be included within the set of variational parameters. While $B=\one_{D}$ represents a fixed choice that is well-suited for the case of periodic boundary conditions, we will see in Section~\ref{s:gauge} that it is beneficial to include one of both boundary vectors $\bm{v}_{\mathrm{L}}$ or $\bm{v}_{\mathrm{R}}$ in the set of variational parameters in the case of open boundary conditions. In order to have a uniform notation, we do not explicitly denote this dependence in the notation for the state $\ket{\Psi[Q,\{R_\alpha\}]}$. Note that it is impossible to absorb the boundary vectors into the matrices $Q(-L/2)$, $R_{\alpha}(-L/2)$ and $Q(L/2)$, $R_{\alpha}(L/2)$ in the case of open boundary conditions. More generally, unlike in the case of generic MPS on finite lattices, it is for cMPS impossible to use a space-dependent bond dimension $D(x)$, since the required continuity of $D$ in combination with its discrete character enforces a constant value.

\section{Computation of expectation values}
\label{s:expectval}
This section is concerned with the computation of expectation values of normally ordered operators. We have already illustrated how to act with annihilation operators and derivatives thereof in the Section~\ref{s:regularity}. With a MPS, the computation of expectation values boils down to a contraction of the physical indices in the network. In the continuum, however, the intuitive notion of physical indices is a bit lost. We therefore start by computing the overlap of two cMPS $\ket{\Psi[Q,\{R_{\alpha}\}]}$, $\ket{\Psi[Q',\{R_{\alpha}'\}]}$, which are given as an expansion in Fock space [Eq.~\eqref{eq:cmpsfockembedding}]. It is clear that the basis states $\hpsid_{\alpha_1}(x_1)\cdots \hpsid_{\alpha_N}(x_N)\ket{\Omega}$ are automatically orthogonal for different $N$, and further that
\begin{multline}
\braket{\Omega|\hpsi_{\beta_N}(y_N)\cdots \hpsi_{\beta_1}(y_1)\hpsid_{\alpha_1}(x_1)\cdots \hpsid_{\alpha_N}(x_N)|\Omega}=\\
\delta_{\alpha_1,\beta_1}\cdots \delta_{\alpha_N,\beta_N} \delta(x_1-y_1)\cdots \delta(x_N-y_N),
\end{multline}
due to the ordering of the arguments $x_1\leq \cdots \leq x_N$ and $y_1\leq \cdots \leq y_N$.
We thus obtain
\begin{multline*}
\braket{\Psi[Q',\{R'_{\alpha}\}]|\Psi[Q,\{R_{\alpha}\}]}=\sum_{N=0}^{+\infty}\sum_{\{\alpha_1,\ldots,\alpha_N\}=1}^{q} \int_{-L/2\leq x_1\leq \cdots \leq x_N\leq +L/2} \d x_1\cdots \d x_N\\
\tr\left[B \Pexp\left(\int_{-L/2}^{x_1} Q(z)\,\d z\right) R_{\alpha_1}(x_1)\cdots R_{\alpha_N}(x_N) \Pexp\left(\int_{x_N}^{+L/2} Q(z)\,\d z\right)\right]\\
\times \tr\left[\overline{B} \Pexp\left(\int_{-L/2}^{x_1} \overline{Q'(z)}\,\d z\right) \overline{R'_{\alpha_1}(x_1)}\cdots \overline{R'_{\alpha_N}(x_N)} \Pexp\left(\int_{x_N}^{+L/2} \overline{Q(z)}\,\d z\right)\right].
\end{multline*}
Using trivial direct product identities such as $\tr[A]\tr[B]=\tr[A\otimes B]$, $(AB)\otimes (CD)=(A\otimes B)(C\otimes D)$ and $\exp(A)\otimes \exp(B)=\exp(A\otimes \one_D+ \one_D \otimes B)$ for $D\times D$ matrices $A$, $B$, $C$ and $D$, the previous expression can be rewritten as
\begin{multline*}
\braket{\Psi[Q',\{R'_{\alpha}\}]|\Psi[Q,\{R_{\alpha}\}]}=\sum_{N=0}^{+\infty}\sum_{\{\alpha_1,\ldots,\alpha_N\}=1}^{q} \int_{-L/2\leq x_1\leq \cdots \leq x_N\leq +L/2} \d x_1\cdots \d x_N\\
\tr\Bigg[(B\otimes \overline{B}) \Pexp\left(\int_{-L/2}^{x_1} [Q(z)\otimes \one_D+\one_D \otimes \overline{Q'(z)}]\,\d z\right) (R_{\alpha_1}(x_1)\otimes \overline{R'_{\alpha_1}(x_1)})\cdots \\
(R_{\alpha_N}(x_N)\otimes \overline{R'_{\alpha_N}(x_N)})\Pexp\left(\int_{x_N}^{+L/2} [Q(z)\otimes \one+\one \otimes \overline{Q'(z)}]\,\d z\right)\Bigg].
\end{multline*}
Reverting the expansion of the path ordered exponential that lead to Eq.~\eqref{eq:cmpsfockembedding}, results in
\begin{multline}
\braket{\Psi[Q',\{R'_{\alpha}\}]|\Psi[Q,\{R_{\alpha}\}]}=\\
\tr\Bigg[(B\otimes \overline{B}) \Pexp\left(\int_{-L/2}^{+L/2} [Q(x)\otimes \one_D+\one_D \otimes \overline{Q'(x)}+\sum_{\alpha=1}^{q} R_{\alpha}(x)\otimes \overline{R'_{\alpha}(x)}]\,\d x\right) \Bigg].
\end{multline}

From the expression above, we can deduce that in the computation of expectation values ($Q'=Q$, $R_\alpha'=R_\alpha$) a central role is played by the local transfer matrix $\voperator{T}(x)$ defined as 
\begin{equation}
\voperator{T}(x)=Q(x)\otimes \one_{D}+\one_{D}\otimes \overline{Q(x)} + \sum_{\alpha=1}^{N} R_{\alpha}(x)\otimes \overline{R_{\alpha}(x)}.\label{eq:transferoperator}
\end{equation}
To this transfer matrix, we can also associate linear maps $\mathscr{T}^{(x)}:\End(\mathbb{C}^{D})\mapsto \End(\mathbb{C}^{D})$ and $\widetilde{\mathscr{T}}^{(x)}:\End(\mathbb{C}^{D})\mapsto \End(\mathbb{C}^{D})$ that map virtual operators $f$ ($D\times D$ matrices) to
\begin{align}
\mathscr{T}^{(x)}(f) &= Q(x) f + f Q(x)^{\dagger}+ \sum_{\alpha=1}^{N} R_{\alpha}(x) f R_{\alpha}(x)^{\dagger},\\
\widetilde{\mathscr{T}}^{(x)}(f) &= f Q(x)  + Q(x)^{\dagger}f+ \sum_{\alpha=1}^{N} R_{\alpha}(x)^{\dagger} f R_{\alpha}(x).
\end{align}

The transfer matrix $\voperator{T}(z)$ is of course strongly related to the transfer matrix $\voperator{E}(n)=\sum_{s} A^s(n)\otimes \overline{A}^s(n)$ that features in expectation values with respect to MPS on the lattice. Indeed, if  $\ket{\Psi[A]}$ is the MPS with matrices $A$ as in Eq.~\eqref{eq:correspondencemps}, then the transfer operator $\voperator{T}(x)$ is related to the transfer operator $\voperator{E}(n)$ of the MPS $\ket{\Psi[A]}$ by $\voperator{E}(n)=\voperator{\one}+a \voperator{T}(na)+\order(a^{2})$. 

The expectation value of any normally ordered operator $\operator{O}=:O[\{\hpsid_{\alpha}\},\{\hpsi_{\beta}\}]:$ can now be computed by first acting with all annihilation operators $\hpsi_{\alpha}(x)$ on the ket $\ket{\Psi[Q,\{R_{\beta}\}]}$ as we did in the Section~\ref{s:regularity}, and similarly acting with the creation operators on the bra. The result of this is the insertion of some operators acting on the virtual system at the corresponding positions, with operators $\operator{U}(x,y)$, $\operator{U}_{\alpha}(x,y)$ or $\operator{U}_{\alpha,\beta}(x,y)$ connecting them. The expectation value is obtained by ``contracting the physical indices'', which results in the inserted virtual operators in the ket combining with those in the bra at the same position\footnote{If there is no insertion at the same position, we can always insert a unit operator $\one_D$}, whereas the contraction of the part in between the local insertions result in a path ordered exponential of the transfer matrix. However, to incorporate the particle statistics, we also need to define generalized transfer operators as
\begin{align}
\voperator{T}_{\alpha}(x)&=Q(x)\otimes \one_{D}+\one_{D}\otimes \overline{Q(x)} + \sum_{\beta=1}^{N} \eta_{\alpha,\beta} R_{\beta}(x)\otimes \overline{R_{\beta}(x)},\\
\voperator{T}_{\alpha,\beta}(x)&=Q(x)\otimes \one_{D}+\one_{D}\otimes \overline{Q(x)} + \sum_{\gamma=1}^{N} \eta_{\alpha,\gamma}\eta_{\beta,\gamma} R_{\gamma}(x)\otimes \overline{R_{\gamma}(x)}.
\end{align}
Note that $\voperator{T}_{\alpha,\alpha}(x)=\voperator{T}(x)$ since $\eta_{\alpha,\beta}^{2}=1$. Given this recipe we can, for example, evaluate the correlation function
\begin{multline}
G^{\alpha,\beta}(x,y)=\braket{\Psi[\overline{Q},\{\overline{R}_{\alpha}\}]|\hpsid_{\alpha}(x)\hpsi_{\beta}(y)|\Psi[Q,\{R_{\alpha}\}]}\\
=\theta(x-y)\tr\bigg[\big(B\otimes \overline{B}\big)\mathscr{P}\ec^{\int_{-L/2}^{+x} \voperator{T}_{\alpha,\beta}(z)\,\d z} \big(R_{\beta}(y)\otimes\one_{D}\big) \mathscr{P}\ec^{\int_{y}^{x}\voperator{T}_{\alpha}(z)\,\d z}\\
\shoveright{\times\big(\one_{D}\otimes \overline{R_{\alpha}(x)}\big)\mathscr{P}\ec^{\int_{x}^{+L/2}\voperator{T}(z)\,\d z}\bigg]\ }\\
+\theta(y-x)\tr\bigg[\big(B\otimes \overline{B}\big)\mathscr{P}\ec^{\int_{-L/2}^{+x} \voperator{T}_{\alpha,\beta}(z)\,\d z} \big(\one_{D}\otimes \overline{R_{\alpha}(x)}\big) \mathscr{P}\ec^{\int_{x}^{y}\voperator{T}_{\beta}(z)\,\d z}\\
\times\big(R_{\beta}(y)\otimes\one_{D}\big)\mathscr{P}\ec^{\int_{y}^{+L/2}\voperator{T}(z)\,\d z}\bigg].\label{eq:corrfungeneric}
\end{multline}
All quantities in this expression, if we could store and manipulate variables with a fully continuous $x$-dependence, are $D^{2}\times D^{2}$ matrices. Since such matrices need to be multiplied, this is an operation with computational complexity of $\order(D^{6})$, or $\order(D^{5})$ if we exploit the tensor-product structure.

For physical systems, we can further simplify Eq.~\eqref{eq:corrfungeneric}. When only bosonic particle species are present, all $\eta_{\alpha,\beta}=1$ and $\voperator{T}=\voperator{T}_{\alpha}=\voperator{T}_{\alpha,\beta}$. In case of the presence of fermionic particle species, we should incorporate the $\mathbb{Z}_{2}$ parity symmetry discussed in the Section~\ref{s:regularity}. We can then define an idempotent parity superoperator $\voperator{P}=P\otimes \overline{P}$ and we obtain $\voperator{P}\voperator{T}\voperator{P}=\voperator{T}$, as well as $\voperator{P}\voperator{T}_{\alpha}\voperator{P}=\voperator{T}_{\alpha}$ and $\voperator{P}\voperator{T}_{\alpha,\beta}\voperator{P}=\voperator{T}_{\alpha,\beta}$. This allows to conclude that $\braket{\Psi[\overline{Q},\{\overline{R}_{\alpha}\}]|\hpsid_{\alpha}(x)\hpsi_{\beta}(y)|\Psi[Q,\{R_{\alpha}\}]}=0$ whenever the particle species $\alpha$ and $\beta$ have different statistics. When $\alpha$ and $\beta$ are both bosonic or both fermionic, it is clear that $\voperator{T}_{\alpha,\beta}=\voperator{T}$ and $\voperator{T}_{\alpha}=\voperator{T}_{\beta}$.

In the case of open boundary conditions, we can define virtual density matrices $l(x),r(x)\in\End(\mathbb{C}^{D})$ which are defined through the initial conditions $l(-L/2)=\bm{v}_{\mathrm{L}}\bm{v}_{\mathrm{L}}^{\dagger}$ and $r(+L/2)=\bm{v}_{\mathrm{R}}\bm{v}_{\mathrm{R}}^{\dagger}$ and the first order differential equations
\begin{align}
\frac{\d\ }{\d x}l(x) &=\widetilde{\mathscr{T}}^{(x)}\big(l(x)\big),&\text{and}&&\frac{\d\ }{\d x}r(x) &=-\mathscr{T}^{(x)}\big(r(x)\big).\label{eq:virtualdensitymatrix}
\end{align}
To these density matrices $l(x)$ and $r(x)$ we associate vectors $\rket{l(x)},\rket{r(x)}\in\mathbb{C}^{D}\otimes\overline{\mathbb{C}^{D}}$ in the ancilla product space. Formally, the solution is given by
\begin{align*}
\rbra{l(x)}&=\rbra{l(-L/2)}\mathscr{P}\ec^{\int_{-L/2}^{x}\voperator{T}(y)\,\d y},\\
\rket{r(x)}&=\mathscr{P}\ec^{\int_{x}^{+L/2}\voperator{T}(y)\,\d y}\rket{r(+L/2)}.
\end{align*}
We can then write 
\begin{align}
\braket{\Psi[\overline{Q},\{\overline{R}_{\alpha}\}]|\Psi[Q,\{R_{\alpha}\}]}&=\Rbraket{l(-L/2)\middle\vert \Pexp\left[\int_{-L/2}^{+L/2} \voperator{T}(x)\,\d x\right]\middle\vert r(+L/2)}\nonumber\\
&=\Rbraket{l(x)|r(x)}=\tr\left[l(x) r(x)\right], \quad \forall x \in \mset{R}.
\end{align}
From the correspondence with completely positive maps, it can be shown that the solution $l(x)$ and $r(x)$ of Eq.~\eqref{eq:virtualdensitymatrix} starting from positive definite initial conditions $l(-L/2)$ and $r(+L/2)$ are positive for any $x\in\mathcal{R}$ (see Theorem~3 in Ref.~\onlinecite{1976CMaPh..48..119L}). The norm is thus guaranteed to be positive. Note that, for the special parameterization of $Q(x)$ in the continuous measurement interpretation [Eq.~\eqref{eq:qunitary}], we can write the determining differential equation for $r(x)$ as
\begin{multline}
\frac{\d\ }{\d x}r(x)=-\mathscr{T}^{(x)}\big(r(x)\big)=\\
-\ic [K(x), r(x)] -\frac{1}{2}\sum_{\alpha=1}^{N} \{R_{\alpha}(x)^{\dagger}R_{\alpha}(x),r(x)\} +\sum_{\alpha=1}^{N}R_{\alpha}(x) r(x) R_{\alpha}(x)^{\dagger}.
\end{multline}
This is a master equation in Lindblad form \cite{1976CMaPh..48..119L} describing the non-equilibrium Markov dynamics of the ancilla (\textit{i.e.} the cavity). Starting from a pure state $r(L/2)=\bm{v}_{\mathrm{R}}\bm{v}_{\mathrm{R}}^{\dagger}$ at $t=-x=-L/2$, it evolves through interaction with the physical system (via the interaction operators $R_{\alpha}$). At a general time $t=-x$, the density matrix $r(x)$ is no longer pure: non-equilibrium evolution is a dissipative process. Note that the evolution is trace preserving, since tracing the equation above results in $\d \tr[r(x)] /\d x=0$. In addition, the corresponding map $\widetilde{\mathscr{T}}^{(x)}$ satisfies $\widetilde{\mathscr{T}}^{(x)}(\one_{D})=0$.

In systems which only contain bosons, all $\eta_{\alpha,\beta}=1$ and there is no need to introduce $\voperator{T}_{\alpha}(x)$, $\voperator{T}_{\alpha,\beta}(x)$, etc. As an alternative to the general recipe described above, we can then also deduce all expectation values of normally ordered operators $\operator{O}=:O[\{\hpsid_{\alpha}\},\{\hpsi_{\alpha}\}]:$ from a generating functional $Z[\{\overline{J}_{\alpha}\},\{J_{\alpha}\}]$ as (see Ref.~\onlinecite{2010PhRvL.105z0401O})
\begin{multline}
\braket{\Psi[\overline{Q},\{\overline{R}_{\alpha}]|:O[\{\hpsid_{\beta}\},\{\hpsi_{\beta}\}]: |\Psi[Q,\{R_{\alpha}\}]}=\\
O\left[\bigg\{\frac{\delta\ }{\delta \overline{J}_{\beta}}\bigg\},\bigg\{\frac{\delta\ }{\delta J_{\beta}}\bigg\}\right]Z[\{\overline{J}_{\alpha}\},\{J_{\alpha}\}]\bigg|_{\overline{J}_{\alpha},J_{\alpha}=0}\label{eq:expecrule}
\end{multline}
with  $\delta\ /\delta J_{\alpha}$ the functional derivative with respect to $J_{\alpha}$, and
\begin{multline}
Z[\{\overline{J}_{\alpha}\},\{J_{\alpha}\}]=\tr\Bigg[\big(B\otimes\overline{B}\big) \Pexp\bigg\{\int_{-L/2}^{+L/2}\d x\,\voperator{T}(x)\\
 + \sum_{\alpha=1}^{N}J_{\alpha}(x)[R_{\alpha}(x)\otimes 1_{D}] +\overline{J}_{\alpha}(x) [1_{D}\otimes \overline{R_{\alpha}(x)}] \bigg\}\Bigg],\label{eq:genfunc}
\end{multline}
which for a system with open boundary conditions results in
\begin{multline}
Z[\{\overline{J}_{\alpha}\},\{J_{\alpha}\}]=\Bigg(l(-L/2)\Bigg\vert\Pexp\bigg\{\int_{-L/2}^{+L/2}\d x\,\voperator{T}(x) \\
 + \sum_{\alpha=1}^{N}J_{\alpha}(x)[R_{\alpha}(x)\otimes 1_{D}] +\overline{J}_{\alpha}(x) [1_{D}\otimes \overline{R_{\alpha}(x)}] \bigg\}\Bigg\vert r(+L/2)\Bigg).\label{eq:genfuncopen}
\end{multline}

Let us now illustrate this approach by defining a generic Hamiltonian for a single-boson system with open boundary conditions\footnote{While we mentioned in Section~\ref{s:bc} that we always assume the matrix functions $Q$ and $R_{\alpha}$ to satisfy the proper boundary conditions, we do not have to use the condition in Eq.~\eqref{eq:qropenbc} at any point in deriving the expectation value of the Hamiltonian $\operator{H}$ in Eq.~\eqref{eq:generichamiltonian}.}
\begin{multline}
\operator{H}=\operator{T}+\operator{V}+\operator{W}=\\
\int_{-L/2}^{+L/2}\d x\,\frac{1}{2m} \left(\frac{\d\ }{\d x} \hpsid(x)\right)\left(\frac{\d\ }{\d x}\hpsi(x)\right)+\int_{-L/2}^{+L/2}\d x\,v(x)\hpsid(x)\hpsi(x)\\
+\frac{1}{2}\int_{-L/2}^{+L/2}\d x\int_{-L/2}^{+L/2}\d y\,w(x,y) \hpsid(x)\hpsid(y)\hpsi(y)\hpsi(x)
\label{eq:generichamiltonian}
\end{multline}
describing particles with mass $m$ that interact with an external potential $v(x)$ and with each other through two-particle interaction $w(x,y)$. 

Using Eq.~\eqref{eq:expecrule} we find (henceforth omitting the arguments $Q$ and $R$ in the state $\ket{\Psi}$)
\begin{equation}
\braket{\Psi|\hpsid(x)\hpsi(x)| \Psi}=\rbraket{l(x)|R(x)\otimes \overline{R}(x)|r(x)},
\end{equation}
and
\begin{multline}
\braket{\Psi|\hpsid(x)\hpsid(y)\hpsi(y) \hpsi(x)| \Psi}=\\
\theta(y-x)\rbraket{l(x)|R(x)\otimes \overline{R(x)} \mathscr{P}\mathrm{e}^{\int_{x}^{y}\d z\, \voperator{T}(z)} R(y)\otimes\overline{R(y)}|r(y)}\\
 +\theta(x-y)\rbraket{l(y)|R(y)\otimes \overline{R(y)} \mathscr{P}\mathrm{e}^{\int_{y}^{x}\d z\, \voperator{T}(z)}R(x)\otimes\overline{R(x)}|r(x)}.
\end{multline}
Defining $R^{(l)}_{x}(x)=R(x)^{\dagger} l(x) R(x)$ for every $x\in[-L/2,+L/2]$ and solving
\begin{align}
\frac{\d\ }{\d y} \rbra{R^{(l)}_{x}(y)}=\rbra{R^{(l)}_{x}(y)}\voperator{T}(y)\label{eq:defrl}
\end{align}
for every $y\in [x,L/2]$, we can write the expectation value of the potential and interaction energy as
\begin{align}
\braket{\Psi|\operator{V}|\Psi}&= \int_{-L/2}^{+L/2}\d x\, v(x) \rbraket{l(x)|R(x)\otimes\overline{R(x)}|r(x)},\\\braket{\Psi|\operator{W}|\Psi}&= \int_{-L/2}^{+L/2}\d x\int_{x}^{+L/2}\d y\, w(x,y) \rbraket{R^{(l)}_{x}(y)|R(y)\otimes\overline{R(y)}|r(y)}.
\end{align}
To evaluate the expectation value of the kinetic energy, we compute
\begin{multline*}
\braket{\Psi|\left(\frac{\d\ }{\d x}\hpsid(x)\right)\left(\frac{\d\ }{\d x}\hpsi(x)\right)|\Psi}=\lim_{x\to y} \frac{\d^{2}\ }{\d x\d y}\braket{\Psi|\hpsid(x)\hpsi(y)|\Psi}\\
\shoveleft{\quad=\lim_{x\to y} \frac{\d^{2}\ }{\d x\d y}\bigg[\theta(y-x)\rbraket{l(x)|(1_{D}\otimes \overline{R(x)})\mathscr{P}\mathrm{e}^{\int_{x}^{y}\d z\, \voperator{T}(z)}(R(y)\otimes 1_{D})|r(y)}}\\
\shoveright{+ \theta(x-y)\rbraket{l(y)|(R(y)\otimes 1_{D})\mathscr{P}\mathrm{e}^{\int_{y}^{x}\d z\, \voperator{T}(z)}(1_{D}\otimes \overline{R(x)})\}|r(x)}\bigg]}\\
\shoveleft{\quad=\lim_{x\to y} \frac{\d\ }{\d x}\Bigg[\theta(y-x)\big(l(x)\big|\big(1_{D}\otimes \overline{R(x)}\big)\mathscr{P}\mathrm{e}^{\int_{x}^{y}\d z\, \voperator{T}(z)}}\\
\shoveright{\times\bigg\{ \big[\voperator{T}(y) ,R(y)\otimes 1_{D}\big] + \big(\d R(y)/\d y \otimes 1_{D}\big) \bigg\}\big\vert r(y)\big)\quad}\\
 \qquad+ \theta(x-y)\big(l(y)\big\vert \bigg\{ \big[\voperator{T},R(y)\otimes 1_{D}\big]+\big(\d R(y)/\d y\otimes 1_{D}\big) \bigg\}\\
\times\mathscr{P}\mathrm{e}^{\int_{y}^{x}\d z\, \voperator{T}(z)}\big(1_{D}\otimes \overline{R(x)}\big)\big|r(x)\big)\Bigg].
\end{multline*}
We have used the defining equations [Eq.~\eqref{eq:virtualdensitymatrix}] in the computation of $\d \rbra{l(y)}/\d y=\rbra{l(y)}\voperator{T}(y)$ and $\d \rket{r(y)}/\d y=-\voperator{T}(y)\rket{r(y)}$. Since $\voperator{T}(y)=Q(y)\otimes 1_{D}+1_{D}\otimes \overline{Q(y)}+R(y)\otimes \overline{R(y)}$, we obtain $[\voperator{T}(y),R(y)\otimes 1_{D}]=[Q(y),R(y)]\otimes 1_{D}$ and thus
\begin{multline*}
\braket{\Psi|\bigg(\frac{\d\ }{\d x}\hpsid(x)\bigg)\bigg(\frac{\d\ }{\d x}\hpsi(x)\bigg)|\Psi}=\\
\shoveleft{\quad\lim_{x\to y} \bigg[\theta(y-x)\big(l(x)\big\vert1_{D}\otimes \big([\overline{Q(x)},\overline{R(x)}]+\d \overline{R(x)}/\d x\big) \mathscr{P}\mathrm{e}^{\int_{x}^{y}\d z\, \voperator{T}(z)}}\\
\shoveright{\times\big( [Q(y) ,R(y)] + \d R(y)/\d y\big) \otimes 1_{D}\big\vert r(y)\big)\quad}\\
+ \theta(x-y)\big(l(y)\big\vert \big( [Q(y),R(y)]+\d R(y)/\d y)\otimes 1_{D}\mathscr{P}\mathrm{e}^{\int_{y}^{x}\d z\, \voperator{T}(z)}\\
\times  1_{D}\otimes \big(1_{D}\otimes[\overline{Q(x)},\overline{R(x)}]+\d \overline{R(x)}/\d x\big)\big\vert r(x)\big)\bigg],
\end{multline*}
where we used the same trick. Note that derivatives with respect to the Heaviside functions (which would produce a diverging contribution $\delta(x-y)$) nicely cancel for both derivatives with respect to $y$ and to $x$. As noted in the Section~\ref{s:regularity}, the regularity condition Eq.~\eqref{eq:regcondition} is automatically fulfilled for the case of a single boson. We thus obtain
\begin{multline}
\braket{\Psi|\operator{T}|\Psi}= \frac{1}{2m}\int_{-L/2}^{+L/2}\d x\, \big(l(x)\big\vert\big([Q(x),R(x)]+\d R(x)/\d x\big)\\
\otimes\big([\overline{Q(x)},\overline{R(x)}]+\d \overline{R(x)}/\d x\big)\big\vert r(x)\big).
\end{multline}
Note that this result could also be obtained by the general strategy outlined at the beginning of this section, \textit{i.e.}\ by acting directly on the cMPS with the operators $\hpsi(x)$ and $\d \hpsi(x) / \d x$ and only afterwards computing the expectation values. However, the generating function approach is very general and relates nicely to the standard approach that is used to compute expectation values in quantum field theory. As for the definition of the state itself, we can also write the generating functional using a path integral, which can be useful for analytic computations or Monte Carlo based evaluation strategies.

\section{Gauge invariance}
\label{s:gauge}
As with a MPS, the map $\Psi$ associating a physical state $\ket{\Psi[Q,\{R_{\alpha}\}]}\in \hilbert_{\mset{R}}^{(\mathrm{F})}$ to the matrix functions $Q:\mset{R}\to \mathbb{C}^{D\times D}$ and $R_{\alpha}:\mset{R}\to\mathbb{C}^{D\times D}$ is not injective, \textit{i.e.} the representation is not unique. For MPS, this so-called \emph{gauge invariance} was rigorously discussed in terms of principal fibre bundles in Ref.~\onlinecite{Haegeman:fk}. Such a rigorous treatment for the case of cMPS is severely complicated by the fact that both the domain and the codomain of the map $\Psi$ are now infinite dimensional. Therefore, it is beyond the scope of the current manuscript, as noted in the introduction. We thus proceed in an intuitive way.

We do expect the existence of a local gauge transformation $g:\mset{R}\to\mathsf{GL}(D,\mathbb{C})$, \textit{i.e.} a position-dependent invertible matrix $g(x)$, that acts on the matrices $Q(x)$ and $R_{\alpha}(x)$ while leaving the physical state $\ket{\Psi[Q,\{R_{\alpha}]}$ invariant. While it is hard to extract the correct transformation formulas for $Q$ and $R_{\alpha}$ from the original cMPS definition in Eq.~\eqref{eq:defcmps}, people with a background in Yang-Mills gauge theories might recognise $Q$ as the connection that generates parallel transport by comparing the $N$-particle wave functions of the Fock space embedding [Eq.~\eqref{eq:cmpsNparticle}] to Wilson lines with insertions of charges transforming according to the adjoint representation, or from recognizing the action of the path integral formulation [Eq.~\eqref{eq:pathintegralrepresentation}] as a Yang-Mills action with a covariant derivative $\frac{\mathrm{d}\ }{\mathrm{d} x} + Q(x)$. The gauge transformation for a cMPS is thus given by
\begin{align}
\tilde{Q}(x)&=g(x)^{-1} Q(x) g(x)+ g(x)^{-1} \frac{\d g}{\d x}(x) ,&\tilde{R}(x)&=g^{-1}(x) R(x) g(x),\label{eq:gaugetransform}
\end{align}
While we prefer the continuum derivation, these transformation formulas can also be obtained by using the correspondence with MPS [Eq.~\eqref{eq:correspondencemps}] and the well-known gauge transformations for MPS \cite{Haegeman:fk}
\begin{align*}
\tilde{A}^{0}(n)&=g((n-1)a)^{-1} A^{0}(n) g(na)\\
&=g((n-1)a)^{-1}g(n a)+a g((n-1)a)^{-1}Q(na)g(na)\\
&=\one_{D}+a\left[-\frac{\d g^{-1}}{\d x}(na) g(n a) + g(na)^{-1} Q(na) g(na)\right]+\order(a^{2}),\\
\tilde{A}^{\alpha}(n) &= g((n-1)a)^{-1} A^{\alpha}(n) g(na)\\
&=\sqrt{a} g(na)^{-1} R_{\alpha}(n a)g(na)+\order(a^{3/2}),\\
\tilde{A}^{(\alpha,\beta)}(n) &= g((n-1)a)^{-1} A^{(\alpha,\beta)}g(na)\\
&=\begin{cases} \frac{a}{2} [ \tilde{R}_{\alpha}(n a) \tilde{R}_{\beta}(n a)+\eta_{\alpha,\beta} \tilde{R}_{\beta}(n a) \tilde{R}_{\alpha}(n a)]+\order(a^{2}),& \alpha\neq \beta\\
\frac{a}{2} \tilde{R}_{\alpha}(n a)^{2}+\order(a^{2}),&\alpha=\beta
\end{cases}\\
&\ldots\nonumber
\end{align*}
Indeed, using $\d g^{-1}(x) /\d x  g(x) = - g^{-1}(x) \d g(x)/ \d x$, we reproduce the transformation formulas of Eq.~\eqref{eq:gaugetransform}. To have an invariant physical state $\ket{\Psi[Q,\{R_{\alpha}\}]}=\ket{\Psi[\tilde{Q},\{R_{\alpha}\}]}$, we also need to transform the boundary matrix as $\tilde{B}=g(L/2)^{-1} B g(-L/2)$. When $B$ is fixed, we need to restrict to gauge transformations that satisfy the boundary condition $g(L/2)^{-1} B g(-L/2)=B$ (\textit{e.g.} $g(L/2)=g(-L/2)$ for $B=\one_D$). In addition, we also require the function $g:\mset{R}\to\mathsf{GL}(D,\mathbb{C})$ to be second order differentiable in order to have new matrix functions  $\tilde{Q}(x)$ and $\tilde{R}_{\alpha}(x)$ which have a well-defined first order derivative. The regularity condition of Eq.~\eqref{eq:regcondition} is not modified by the gauge transformation and puts no further constraints on the set of allowed gauge transformations. Since this condition follows from physical considerations which are left invariant by gauge transformations, it would be strange if we obtained a different result.

As for MPS, we can use the gauge fixing conditions to impose a certain canonical form on the matrices $Q(x)$ and $R_{\alpha}(x)$. Suppose we want to impose a gauge fixing condition such that $\tilde{Q}(x)$ is of the form in Eq.~\eqref{eq:qunitary}, corresponding to the cMPS construction from continuous measurement. It is equivalent to the \emph{left orthonormalization condition} of MPS and boils down to imposing
\begin{displaymath}
\tilde{Q}(x)+\tilde{Q}(x)^\dagger +\sum_{\alpha=1}^{q} \tilde{R}_{\alpha}(x)^\dagger \tilde{R}_{\alpha}(x)=0
\end{displaymath}
for every $x\in\mathcal{R}$. Inserting the explicit form of $\tilde{Q}(x)$ and $\tilde{R}_{\alpha}(x)$ in terms of the original $Q(x)$, $R_{\alpha}(x)$ and $g(x)$ [Eq.~\eqref{eq:gaugetransform}], we obtain that $g(x)$ should be a solution of the differential equation
\begin{multline}
\begin{split}
\frac{\d\ }{\d x} \left[ \left(g^{-1}(x)\right)^\dagger g^{-1}(x)\right]&= \left(g^{-1}(x)\right)^\dagger g^{-1}(x) Q(x) + Q(x)^\dagger \left(g^{-1}(x)\right)^\dagger g^{-1}(x)\\
&\qquad\qquad+\sum_{\alpha=1}^{q} R_{\alpha}(x)^\dagger \left(g^{-1}(x)\right)^\dagger g^{-1}(x) R_{\alpha}(x)\\
&=\tilde{\mathscr{T}}^{(x)}\left[\left(g^{-1}(x)\right)^\dagger g^{-1}(x)\right].
\end{split}
\end{multline}
Clearly, this differential equation only determines $g(x)$ up to a unitary prefactor. Put differently, for any solution $g(x)$ of this equation, $g'(x)=u(x) g(x)$ with $u(x)$ a unitary matrix is an equally valid solution. We can use the remaining gauge freedom $u(x)\in\mathsf{U}(D)$ to diagonalize $r(x)$ at every point $x$, hence obtaining the \emph{left-canonical form}.

However, at this point it becomes important to discuss the boundary conditions that should be satisfied by solutions $g(x)$. If the boundary matrix $B$ is fixed, we need to impose $g^{-1}(+L/2) B g(-L/2)=B$. This is a highly non-trivial condition and it is not certain that such solutions exist. For periodic boundary conditions with $B=\one_{D}$, it logically results in $g(+L/2)=g(-L/2)$. Translation-invariant states with periodic boundary conditions can be subjected to the the same treatment as the translation-invariant states in the thermodynamic limit, which are discussed in the next section. Henceforth, we restrict to the case of open boundary conditions with $B=\bm{v}_{\mathrm{R}}\bm{v}_{\mathrm{L}}^{\dagger}$. From this, we can derive the conditions
\begin{align*}
\bm{v}_{\mathrm{L}}^{\dagger} g(-L/2) &= \alpha \bm{v}_{\mathrm{L}}^{\dagger} &g^{-1}(+L/2)\bm{v}_{\mathrm{R}}&=\frac{1}{\alpha} \bm{v}_{\mathrm{R}}
\end{align*}
for some non-zero $\alpha\in\mathbb{C}$. However, we can easily fix $\alpha=1$ by substituting $g(x)\leftarrow g'(x)=g(x)/\alpha$, since the constant gauge transformation $\alpha \one_{D}$ acts trivially on $Q$ and $R$, \textit{i.e.} it is within the kernel of the gauge group action. Nevertheless, the resulting boundary conditions are still highly non-trivial and it is not assured by the standard theory of differential equations that there exist solutions satisfying both conditions simultaneously. Hence, it is better to restrict to a single boundary condition such as $g(-L/2)=\one_{D}$ and do not impose any condition on $g(+L/2)$. The value of $g(+L/2)$ is then completely determined by the differential equation (up to the unitary prefactor). Consequently, we then also have to transform the right boundary vector as $\tilde{\bm{v}}_{\mathrm{R}}=g^{-1}(+L/2) \bm{v}_{\mathrm{R}}$. This implies that $\bm{v}_{\mathrm{R}}$ is part of the variational degrees of freedom, and should also be included in \textit{e.g.} the variational optimization for finding ground states. Note that the boundary conditions for $g(x)$ are inherently imposed by the representation of the state, and are not related to or influenced by the physical conditions that need to be satisfied by $Q$ and $R$, as discussed in Section~\ref{s:bc}.

Alternatively, we can also impose the \emph{right orthonormalization condition}, which boils down to
\begin{equation}
\tilde{Q}(x)+\tilde{Q}(x)^{\dagger}+\sum_{\alpha=1}^{N}\tilde{R}_{\alpha}(x)\tilde{R}_{\alpha}(x)^{\dagger}=0
\end{equation}
and implies that
\begin{equation}
\tilde{Q}(x)=-\ic K(x) -\frac{1}{2}\sum_{\alpha=1}^{N}\tilde{R}_{\alpha}(x)\tilde{R}_{\alpha}(x)^{\dagger}
\end{equation}
with $K(x)$ a Hermitian matrix. Starting from an arbitrary cMPS with matrices $Q(x)$ and $R_{\alpha}(x)$, we obtain new matrices $\tilde{Q}(x)$ and $\tilde{R}_{\alpha(x)}$ according to Eq.~\eqref{eq:gaugetransform}, which satisfy the above relations if $g(x)$ is a solution of
\begin{multline}
\begin{split}
\frac{\d\ }{\d x} \left[ g(x) g(x)^{\dagger} \right]&= -Q(x) g(x) g(x)^{\dagger} - g(x) g(x)^{\dagger} Q(x)^\dagger-\sum_{\alpha=1}^{q} R_{\alpha}(x)g(x)g(x)^{\dagger}R_{\alpha}(x)^{\dagger} \\
&=-\mathscr{T}^{(x)}\left[g(x) g(x)^\dagger\right].
\end{split}
\end{multline}
Clearly, for any solution $g(x)$, we obtain a family of solutions $g'(x)=g(x) u(x)$ with $u(x)\in\mathsf{U}(D)$. This unitary freedom can be fixed by diagonalizing $l(x)$, resulting in the \emph{right-canonical form}. As for the left-canonical form, one has to pay careful attention to the boundary conditions that need to be satisfied by $g$. For a system with open boundary conditions, the easiest solution is again to include one of the boundary vectors in the set of the variational parameters and also transform it under the action of the gauge transform. 

Note that we can also define a gauge transformation $g(x)$ for the cMPS $\ket{\Psi[Q,\{R_{\alpha}\}]}\in\varM_{\text{cMPS}}$ so that
\begin{equation}
\tilde{Q}(x)=g(x)^{-1} Q(x) g(x)+g(x)^{-1} \frac{\d g}{\d x}(x)=0.
\end{equation}
It is sufficient to choose
\begin{equation}
g(x)=\mathscr{P}\!\exp\left[\int^{+L/2}_{x} Q(y)\,\d y\right] g_0
\end{equation}
with $g_{0}$ some arbitrary integration factor that is fixed by the boundary conditions. For example, if we require $g(-L/2)=\one_{D}$ then $g_0=\left(\mathscr{P}\!\exp\left[\int^{+L/2}_{-L/2} Q(y)\,\d y\right]\right)^{-1}$ and we also need to transform $\bm{v}_{\mathrm{R}}\leftarrow \bm{\tilde{v}}_{\mathrm{R}}= g(+L/2)^{-1}\bm{v}_{\mathrm{R}}=g_0^{-1}\bm{v}_{\mathrm{R}}$. Hence, the cMPS can now be written as
\begin{equation}
\ket{\Psi[\{\tilde{R}_{\alpha}\}]}=\bm{v}_{\mathrm{L}}^{\dagger} \mathscr{P}\!\exp\left[\int_{-L/2}^{+L/2}\d x\, \sum_{\alpha=1}^{N}\tilde{R}_{\alpha}(x) \otimes \hpsid_{\alpha}(x) \right]\bm{\tilde{v}}_{\mathrm{R}}\ket{\Omega}.\label{eq:formulationlinkwithmeanfield}
\end{equation}
This formulation is close in spirit to the bosonic mean field ansatz
\begin{displaymath}
\ket{\varphi}=\exp\left(\int_{-L/2}^{+L/2}\varphi(x) \hpsid(x)\,\d x\right)\ket{\Omega}
\end{displaymath}
with $\varphi$ a scalar (complex-valued) function, since it identifies the mean field ansatz with a cMPS with bond dimension $D=1$. This mean field ansatz lies at the basis of the Gross-Pitaevskii equation \cite{Gross:1961aa,Pitaevskii:1961aa}, that is still used today with great success. All variational degrees of freedom are now contained in the matrices $\tilde{R}_{\alpha}(x)$ (and $\bm{\tilde{v}}_{\mathrm{R}}$), and all gauge degrees of freedom have been eliminated. However, 
we do not employ this particular choice of gauge in the remainder of this manuscript as it also has some downsides. For example, translation-invariant states $\ket{\Psi[Q,R_{\alpha}]}$ can be obtained by choosing the matrices $Q$ and $R_{\alpha}$ $x$-independent (see next subsection). However, this particular gauge transformation maps the $x$-independent matrices $R_{\alpha}$ to $x$-dependent matrices $\tilde{R}_{\alpha}(x)=\ec^{+Q x} R_{\alpha}\ec^{-Q x}$, so that translation invariance is less easily recognized.

\section{Translation invariance and the thermodynamic limit}
\label{s:ti}
When using cMPS to approximate ground states of translation invariant Hamiltonians, we can restrict to the subclass of uniform cMPS $\ket{\Psi(Q,\{R_{\alpha}\})}$, which are obtained from taking $Q(x)=Q$ and $R_{\alpha}(x)=R_{\alpha}$ constant $x$-independent $D\times D$ matrices in $\ket{\Psi[Q,\{R_{\alpha}\}]}$. This approach is valid both for a finite system with periodic boundary conditions ($B=\one_{D}$) or for a system in the thermodynamic limit ($\lvert\mset{R}\rvert=L\to \infty$ or thus $\mset{R}\to \mathbb{R}$), where the precise value of the boundary matrix $B$ should be irrelevant and should not appear in any normalised expectation value. We henceforth restrict to the latter case. The transfer operator $\voperator{T}=Q\otimes 1_{D}+1_{D}\otimes\overline{Q}+\sum_{\alpha=1}^{q} R_{\alpha}\otimes\overline{R}_{\alpha}$ also becomes translation invariant and $\Pexp[\int_{y}^{z}\d x\, \voperator{T}]=\exp[\voperator{T}(z-y)]$. The normalization of the state $\ket{\Psi(Q,R)}$ is given by $\lim_{L\to\infty}\tr\big[(B\otimes\overline{B})\exp(\voperator{T} L)\big]$. If $\mu=\max_{\lambda\in\sigma(\voperator{T})}\{\Re(\lambda)\}$, where $\sigma(\voperator{T})$ denotes the spectrum of $\voperator{T}$ and $\Re$ the real part, then $\braket{\Psi(\overline{Q},\{\overline{R}_{\alpha}\})|\Psi(Q,\{R_{\alpha}\})}\sim \lim_{L\to\infty} \exp(\mu L)$. Normalizing this state by multiplying it with $\exp(-\mu L)$ results in $Q\leftarrow Q-\mu/2 \one_{D}$ and $\voperator{T}\leftarrow \voperator{T}-\mu \voperator{\one}$, so that the new transfer operator $\voperator{T}$ has at least one eigenvalue for which the real part is zero and no eigenvalue has a positive real part. Let us assume that the eigenvalue $\lambda$ with $\Re \lambda=0$ is unique. If $\rket{r}$ is the corresponding right eigenvector, then we can write the eigenvalue equation as $\mathscr{T}(r)=\lambda r$ with $r$ the associated virtual density matrix. Hermitian conjugation learns that $\mathscr{T}(r^{\dagger})=\overline{\lambda} r^{\dagger}$, so that the uniqueness of the eigenvalue with $\Re \lambda=0$ implies that $\lambda=\overline{\lambda}=0$ and $r^{\dagger}=\ec^{\ic \phi} r$, where we can choose the phase of the eigenvector so that $r$ is Hermitian. Similarly, the virtual density matrix $l$ associated to the left eigenvector $\rket{l}$ can also be chosen Hermitian. 

Having a unique eigenvalue zero and $\Re(\lambda)<0$ for all other eigenvalues $\lambda$ corresponds to the generic case, as can be better appreciated by referring to the well-known results for MPS\cite{1992CMaPh.144..443F,2006quant.ph..8197P,Haegeman:fk}. Indeed, a full categorisation of the eigenvalue structure of $\voperator{T}$ can be obtained by identifying\footnote{While we take a standard matrix logarithm, it also makes sense to define the linear maps $\mathscr{T}$, $\tilde{\mathscr{T}}$ as the logarithm of ---or the generator for--- the completely positive maps $\mathscr{E}$ and $\tilde{\mathscr{E}}$ associated to the left or right action of $\voperator{E}$. However, not all completely positive maps have a natural logarithm associated to it, as was shown in Ref.~\onlinecite{2008CMaPh.279..147W}.}
\begin{equation}
\voperator{T}=\lim_{a\to 0} \frac{1}{a} \ln \voperator{E}
\end{equation}
with $\voperator{E}$ the corresponding transfer operator of the uniform MPS $\ket{\Psi(A)}$ with $A$ related to $Q$ and $R_{\alpha}$ as in Eq.~\eqref{eq:correspondencemps}. The set of MPS with a well-defined thermodynamic limit correspond to the injective or pure MPS, for which the transfer operator $\voperator{E}$ has a single eigenvalue $1$ that maps to the eigenvalue zero of $\voperator{T}$. The corresponding left and right eigenvectors $\rbra{l}$ and $\rket{r}$ correspond to strictly positive Hermitian operators $l$ and $r$ (\textit{i.e.}\ they have full rank). All other eigenvalues of $\voperator{E}$ lie strictly within the unit circle and map to eigenvalues of $\voperator{T}$ with strictly negative real part. If the left and right eigenvectors corresponding to eigenvalue $0$ are normalized such that $\rbraket{l|r}=1$, then $\lim_{L\to\infty} \exp(\voperator{T} L)=\rket{r}\rbra{l}$ and we obtain
\begin{equation}
\braket{\Psi(\overline{Q},\{\overline{R}_{\alpha}\})|\Psi(Q,\{R_{\alpha}\})}=\rbraket{l|B\otimes \overline{B}|r}.
\end{equation}
In expectation values of local operators, this overall factor always appears, but the rest of the expression will not depend on $B$. Hence, the $B$-dependence is cancelled by considering normalized expectation values, or by henceforth choosing $B$ such that $\braket{\Psi(Q,\{R_{\alpha}\})|\Psi(Q,\{R_{\alpha}\})}=\rbraket{l|B\otimes \overline{B}|r}=1$. 

For uniform cMPS, the gauge invariance is restricted to global transformations $Q\leftarrow\tilde{Q}=g Q g^{-1}$ and $R_{\alpha}\leftarrow \tilde{R}_{\alpha}=g R_{\alpha} g^{-1}$ with $g\in\group{GL}(\mathbb{C},D)$. This gauge transformation can be used to impose the left or right orthonormalization conditions. Left orthonormalization boils down to fixing the left eigenvector $l$ of eigenvalue $0$ to $l=\one_{D}$, which results in $Q=-\ic K-1/2 \sum_{\alpha=1}^{q} R_{\alpha}^{\dagger}R_{\alpha}$ with $K$ a Hermitian matrix. The remaining unitary gauge freedom can be used to diagonalize $r$, bringing $Q$ and $R_{\alpha}$ in the left-canonical form. The right-canonical form is obtained analogously. In principle, an exact computation of the left and right eigenvectors $l$ and $r$ corresponding to the eigenvalue with largest real part $\lambda$ of the transfer operator $\voperator{T}$ are computationally costly operations [$\order(D^{6})$]. By using an explicit parameterization of the left-canonical form in terms of $R_{\alpha}$ and the Hermitian matrix $K$, we know exactly that $\lambda=0$ and $l=\one_{D}$. It is then possible to obtain $r$ with an iterative solver with computational efficiency $\order(D^{3})$. 

By imposing the physical requirements discussed at the end of Section~\ref{s:regularity}, we can define the parity superoperator $\voperator{P}$ as in Section~\ref{s:expectval}. Since $\voperator{P}\voperator{T}\voperator{P}=\voperator{T}$, we can expect that the left and right eigenvectors $\rket{l}$ and $\rket{r}$ corresponding to the zero eigenvalue satisfy $\rbra{l}\voperator{P}=\rbra{l}$ and $\voperator{P}\rket{r}=\rket{r}$, or thus $P^{\dagger} l P = l$ and $P r P^{\dagger}=r$. Note that we can always choose the gauge such that $P$ is Hermitian. In addition, it is easy to prove that $\voperator{T}_{\alpha}$ also has an eigenvalue zero even if $\alpha$ refers to a fermionic particle species so that $\voperator{T}_{\alpha}\neq \voperator{T}$. The corresponding left and right eigenvectors are in that case given by $l_{\alpha}=l P=P^{\dagger} l$ and $r_{\alpha} =P r=r P^{\dagger}$, whereas they equal $l$ and $r$ if $\alpha$ is a bosonic particle.

We can now evaluate correlation functions as
\begin{multline}
C_{\alpha,\beta}(x,y)=\braket{\Psi(\overline{Q},\{\overline{R}_{\alpha}\})|\hpsid_{\alpha}(x)\hpsi_{\beta}(y)|\Psi(Q,\{R_{\alpha}\})}\\
=\theta(x-y)\rbraket{l|[R_{\beta}\otimes\one_{D}]\ec^{\voperator{T}_{\alpha}(x-y)}[\one_{D}\otimes \overline{R_{\alpha}}]|r}\\
+\theta(y-x)\rbraket{l|[\one_{D}\otimes \overline{R_{\alpha}}]\ec^{\voperator{T}_{\alpha}(y-x)}[R_{\beta}\otimes\one_{D}]|r},\label{eq:corrfunti}
\end{multline}
where we have used the physical requirement $\voperator{T}_{\alpha,\beta}=\voperator{T}$ and $\voperator{T}_{\alpha}=\voperator{T}_{\beta}$ for non-vanishing correlation functions (see Section~\ref{s:expectval}). The correlation function $C_{\alpha,\beta}(x,y)$ is translation invariant and we define $C_{\alpha,\beta}(x,y)=C_{\alpha,\beta}(y-x)$. When $\alpha$ is bosonic and $\beta$ fermionic, we automatically have $C_{\alpha,\beta}(x)=0$ if the parity considerations from Section~\ref{s:regularity} are correctly built in. In the long-range limit, we obtain $\lim_{\lvert x\rvert \to \infty}C_{\alpha,\beta}(x)=\rbraket{l|R_{\beta}\otimes\one_{D}|r_{\alpha}}\rbraket{l_{\alpha}|\one_{D}\otimes \overline{R_{\alpha}}|r}$. When both $\alpha$ and $\beta$ refer to fermionic particle species, this limiting value is automatically zero (also under the assumption that parity is correctly built into the matrices). When both indices refer to bosonic particles, a non-zero value is possible in the case of Bose-Einstein condensation. We should then define a connected correlation function $\tilde{C}_{\alpha,\beta}(x)$, which decays exponentially as $\lim_{\lvert x \rvert\to \infty}\tilde{C}_{\alpha,\beta}(x)=\order(\exp[-\lvert x\rvert/\xi_{\text{c}}])$ with $\xi_{c}=(\Re \lambda_{1})^{-1}$, where $\lambda_{1}$ is the eigenvalue of $\voperator{T}_{\alpha}$ with second largest real part (\textit{i.e.} skipping eigenvalue $\lambda_{0}=0$). Clearly, $C_{\alpha,\beta}(x)$ is continuous at $x=0$. We can then compute the first derivative, which is only continuous at $x=0$ if we impose the regularity conditions in Eq.~\eqref{eq:regcondition}. This is another way to derive these conditions. If Eq.~\eqref{eq:regcondition} is satisfied, then the second derivative of $C_{\alpha,\beta}(x)$ at $x=0$ (which gives the expectation value of the kinetic energy density $\operator{t}$ up to a factor $-1/2m$) is finite and automatically continuous. The third derivative is then finite but will not be continuous in general, without imposing further conditions as discussed in Appendix~\ref{a:higherorderregularity}.

We define the Fourier transformed correlation function
\begin{equation}
n_{\alpha,\beta}(p,p')=\int_{-\infty}^{+\infty} \frac{\d x}{2\pi} \int_{-\infty}^{+\infty}\frac{\d y}{2\pi}\, C_{\alpha,\beta}(x,y)\ec^{\ic p x - \ic p' y}= \delta (p'-p) n_{\alpha,\beta}(p)
\end{equation}
with 
\begin{equation}
n_{\alpha,\beta}(p)=\int_{-\infty}^{+\infty} \frac{\d x}{2\pi} C_{\alpha,\beta}(x) \ec^{-\ic p x}.
\end{equation}
In order to evaluate $n_{\alpha,\beta}(p)$, it is important to separate $\exp(\voperator{T}_{\alpha}x)$ into two parts. The first part is given by $\voperator{S}_{\alpha}=\rket{r_{\alpha}}\rbra{l_{\alpha}}$, the projector onto the eigenspace corresponding to eigenvalue $0$ of $\voperator{T}_{\alpha}$, and yields a singular contribution to the integral. If we define the complementary projector $\voperator{Q}_{\alpha}=\one-\voperator{S}_{\alpha}$, then the remaining part
\begin{displaymath}
\exp(\voperator{T}_{\alpha}x)-\voperator{S}_{\alpha}=\voperator{Q}_{\alpha}\exp(\voperator{T}_{\alpha}x) \voperator{Q}_{\alpha}=\voperator{Q}_{\alpha}\exp(\voperator{Q}_{\alpha}\voperator{T}_{\alpha}\voperator{Q}_{\alpha}x) \voperator{Q}_{\alpha}\label{eq:singulardecompositionT}
\end{displaymath}
is well behaved in the Fourier transform, since all of its eigenvalues decay exponentially $x$. If we then introduce the notation $\voperator{Q}_{\alpha}(-\voperator{T}_{\alpha}\pm\ic p)^{-1}\voperator{Q}_{\alpha}=(-\voperator{T}_{\alpha}\pm\ic p)^{\mathsf{P}}$, which is well defined even at $p=0$ because the zero eigensector of $\voperator{T}_{\alpha}$ is projected out, we can rewrite $n_{\alpha,\beta}(p)$ as
\begin{multline}
n_{\alpha,\beta}(p)=2\pi \delta(p) \rbraket{l|\one_{D}\otimes \overline{R_{\alpha}}|r_{\alpha}}\rbraket{l_{\alpha}|R_{\beta}\otimes\one_{D}|r}\\
+\rbraket{l|[\one_{D}\otimes \overline{R_{\alpha}}] (-\voperator{T}_{\alpha}+\ic p)^{\mathsf{P}} [R_{\beta}\otimes\one_{D}]|r}\\
+\rbraket{l|[R_{\beta}\otimes\one_{D}] (-\voperator{T}_{\alpha}-\ic p)^{\mathsf{P}} [\one_{D}\otimes \overline{R_{\alpha}}]|r}. \label{eq:cmpsmomentumoccupation}
\end{multline}
The first term is only present for bosonic particles that have condensed. It would also disappear in the Fourier transformation of the connected correlation function $\tilde{C}(x,y)$. If we define Fourier transformed field operators $\hPsi(p)$ ---no confusion between the state $\ket{\Psi}$ and the momentum-space operator $\hPsi$ should arise--- as
\begin{equation}
\hPsi(p)=\frac{1}{\sqrt{2\pi}}\int_{-\infty}^{+\infty}\d x\,\hpsi(x)\ec^{-\ic p x},
\end{equation}
then it is easy to see why we have used the suggestive notation $n_{\alpha,\beta}$ for the Fourier transform of $C_{\alpha,\beta}$. We obtain
\begin{equation}
\braket{\Psi(\overline{Q},\{\overline{R}_{\alpha}\})|\hPsid_{\alpha}(p)\hPsi_{\beta}(p')|\Psi(Q,\{R_{\alpha}\})}=\delta(p-p')n_{\alpha,\beta}(p).\label{eq:defmomoccnum}
\end{equation}
Hence, $n_{\alpha,\beta}(p)$ describes the occupation number of momentum levels. The large-$p$ behavior of $n_{\alpha,\beta}(p)$ follows from the regularity of $C_{\alpha,\beta}(x)$. At first sight, Eq.~\eqref{eq:cmpsmomentumoccupation} might seem to decay as $\order(p^{-1})$. However, if the regularity conditions in Eq.~\eqref{eq:regcondition} are satisfied, then the momentum occupation number $n_{\alpha,\beta}(p)$ has to decay as $\order(p^{-4})$ for large values of $p$. We can show this explicitly. For $\lvert p\rvert$ larger than the eigenvalue of $\voperator{T}_{\alpha}$ with the largest absolute value, we can expand $(-\voperator{T}_{\alpha}\pm \ic p)^{\mathsf{P}}$ as
\begin{equation}
(-\voperator{T}_{\alpha}\pm \ic p)^{\mathsf{P}}=\mp \ic\frac{\voperator{Q}_{\alpha}}{p}\sum_{n=0}^{+\infty} \left(\pm \ic \frac{\voperator{T}_{\alpha}}{p}\right)^n=\mp \ic \frac{\voperator{Q}_{\alpha}}{p} +\frac{\voperator{T}_{\alpha}}{p^2}\pm \ic \frac{\voperator{T}_{\alpha}^2}{p^3}-\frac{\voperator{T}_{\alpha}^3}{p^4}+\order(p^{-5}).
\end{equation}
We now have to show that by plugging this expansion into Eq.~\eqref{eq:cmpsmomentumoccupation}, the first three terms vanish. The first term is trivial, if particle type $\alpha$ is bosonic so that $\voperator{Q}_{\alpha}=\voperator{\one}-\rket{r}\rbra{l}$. For the fermionic case, one has to employ the parity conservation. Using the regularity conditions of Eq.~\eqref{eq:regcondition} and $\eta_{\alpha,\gamma}=\eta_{\beta,\gamma}$ for non-vanishing correlation functions ---$\alpha$ and $\beta$ are of both bosonic or both fermionic--- we can show that
\begin{align*}
\voperator{T}_{\alpha} [R_{\beta}\otimes \one_{D}]\rket{r}=[R_{\beta}\otimes\one_{D}]\voperator{T}\rket{r}+[Q,R_{\beta}]\otimes\one_{D}\rket{r}=[Q,R_{\beta}]\otimes\one_{D}\rket{r}
\end{align*}
and similarly
\begin{align*}
\voperator{T}_{\alpha} [\one_{D}\otimes \overline{R_{\alpha}}]\rket{r}&=\one_{D}\otimes[\overline{Q},\overline{R_{\alpha}}]\rket{r},\\
\rbra{l}[R_{\beta}\otimes\one_{D}]\voperator{T}_{\alpha} &=\rbra{l}[R_{\beta},Q]\otimes\one_{D},\\
\rbra{l}[\one_{D}\otimes \overline{R_{\alpha}}]\voperator{T}_{\alpha} &=\rbra{l}\one_{D}\otimes [\overline{R_{\alpha}},\overline{Q}].
\end{align*}
These results can be used to show that both the second and third term in the expansion vanish when they are plugged into Eq.~\eqref{eq:cmpsmomentumoccupation}. The first non-vanishing term is thus of order $p^{-4}$. Because $n_{\alpha,\beta}(p)$ is a dimensionless quantity, this asymptotic behavior allows us to introduce a momentum cutoff $\Lambda$ as
\begin{equation}
\Lambda^4=\lim_{p\to\infty} \lvert p^4 n_{\alpha,\beta}(p)\rvert=\lvert \rbraket{l|[\one_{D}\otimes \overline{R_{\alpha}}] \voperator{T}_{\alpha}^3 [R_{\beta}\otimes\one_{D}]|r}+\rbraket{l|[R_{\beta}\otimes\one_{D}] \voperator{T}_{\alpha}^3 [\one_{D}\otimes \overline{R_{\alpha}}]|r}\rvert, 
\end{equation}
where the absolute value is not required if we use $\beta=\alpha$. The eigenvalue spectrum of $\voperator{T}_{\alpha}$ thus provides a definition for an ultraviolet cutoff scale $a=\Lambda^{-1}$. Rather than defining the ultraviolet cutoff scale $a=\Lambda^{-1}$ through the total particle density
\begin{equation}
\rho_{\alpha,\beta}=\int_{-\infty}^{+\infty}\frac{\d p}{2\pi}\, n_{\alpha,\beta}(p),
\end{equation}
we have now defined a UV cutoff scale $\Lambda$ based on the large momentum behavior of the momentum occupation number $n_{\alpha,\beta}(p)$.

For two pure uniform cMPS $\ket{\Psi(Q,\{R_{\alpha}\})}$ and $\ket{\Psi(Q',\{R_{\alpha}'\})}$ we can define a superoperator $\voperator{T}_{\text{mixed}}=Q'\otimes\one_{D}+\one_{D}\otimes \overline{Q}+\sum_{\alpha=1}^{N}R_{\alpha}'\otimes \overline{R_{\alpha}}$ so that the $\braket{\Psi(Q,\{R_{\alpha}\})|\Psi(Q',\{R_{\alpha}'\})}$ decays as $\lim_{L\to+\infty}\exp(\lambda L)$, with $\lambda$ the eigenvalue with largest real part of $\voperator{T}_{\text{mixed}}$. If the two uniform cMPS are inequivalent, $\Re(\lambda) < 0$ and there is an infrared orthogonality catastrophe. If $\Re(\lambda)=0$, then we can define a phase $\phi=\Im(\lambda)$ and a gauge transformation $g\in\mathsf{GL}(D;\mathbb{C})$ such that $Q'=g Q g^{-1} +\ic \phi$ and $R'_{\alpha}=g R_{\alpha} g^{-1}$. With $f$ being the right eigenvector corresponding to eigenvalue $\lambda=\ic \phi$ of $\voperator{T}_{\text{mixed}}$, $g$ can be obtained as $g=f r^{-1}$.

Let us also illustrate how to compute the expectation value of a translation invariant Hamiltonian. The generic Hamiltonian in Eq.~\eqref{eq:generichamiltonian} becomes translation invariant for $v(x)=v$ and $w(x,y)=w(y-x)$ with $w(x)=w(-x)$. Since the uniform cMPS is extensive, expectation values are proportional to the volume and it makes more sense to compute the expectation values of the kinetic, potential and interaction energy densities $\operator{t}$, $\operator{v}$ and $\operator{w}$. We obtain
\begin{align}
\braket{\Psi(\overline{Q},\{\overline{R}_{\alpha}\})|\operator{t}|\Psi(Q,\{R_{\alpha}\})}&=\frac{1}{2m}\rbraket{l|[Q,R]\otimes [\overline{Q},\overline{R}]|r},\\
\braket{\Psi(\overline{Q},\{\overline{R}_{\alpha}\})|\operator{v}|\Psi(Q,\{R_{\alpha}\})}&=v\rbraket{l|R\otimes \overline{R}|r},
\end{align}
\begin{align}
\braket{\Psi(\overline{Q},\{\overline{R}_{\alpha}\})|\operator{w}|\Psi(Q,\{R_{\alpha}\})}&=\int_{0}^{+\infty}\d z\,w(z)\rbraket{l|R\otimes \overline{R} \mathrm{e}^{\voperator{T} z} R\otimes\overline{R}|r}.
\end{align}
If $w(z)$ has a Laplace transform $\mathscr{L}[w](\sigma)=\int_{0}^{+\infty}\d z w(z) \exp(-\sigma z)$ that is defined for $\Re \sigma \geq 0$, we obtain
\begin{align}
\braket{\Psi|\operator{w}|\Psi}&=\rbraket{l|R\otimes \overline{R}\ \mathscr{L}[w](-\voperator{T}) R\otimes\overline{R}|r}.
\end{align}
Note that translation invariance has allowed the parameterization of a field with a continuous number of degrees of freedom by a discrete number of degrees of freedom. Having $l$ and $r$, the computational cost is $\order(D^{6})$ when long-range interactions are present, since we then have to compute an arbitrary function $\mathscr{L}[w]$ of the transfer operator $\voperator{T}$, unless $w$ is such that there is an exact or approximate (iterative) strategy for evaluating the action of $\mathscr{L}[w](-\voperator{T})$ on a vector efficiently. One particular example is the case of strictly local interactions $w(x-y)\sim \delta(x-y)$. The interaction energy (density) can then be computed with computational complexity of $\order(D^{3})$ just like the potential and the kinetic energy density.

\section{Tangent vectors of continuous matrix product states}
\label{s:tangent}
\subsection{Generic case}
For MPS, a new algorithm for time evolution and variational optimization (via imaginary time evolution) was recently constructed using the time-dependent variational principle\cite{2011arXiv1103.0936H}. An essential ingredient of this algorithm is the study of (infinitesimally) small variations of MPS, \textit{i.e.} the set of MPS tangent vectors. Indeed, it was rigorously proven that the set of MPS can be given the structure of a variational manifold with a well-defined tangent space\cite{Haegeman:fk} by eliminating some singular points or regions. While we do expect the same theorems to hold for cMPS, the infinite dimensionality of the parameter space and Hilbert space might require a different proof strategy, especially in the absence of translation invariance. As noted several times before, this would be beyond the scope of this paper.

Given the practical use of tangent vectors, we nevertheless proceed, albeit in a more intuitive manner. Let us assume that we do have an open subset of cMPS with fixed bond dimension $D$ that constitute a (complex) manifold $\mset{M}_{\mathrm{cMPS}}\subset \hilbert$. At any base point $\ket{\Psi[Q,\{R_{\alpha}\}]}\in \mset{M}_{\mathrm{cMPS}}$, we can construct a (holomorphic) tangent space $T_{\ket{\Psi[Q,\{R_{\alpha}\}]}} \mset{M}_{\mathrm{cMPS}} \subset \hilbert$. If the collective index $i=1,\ldots,D^2$ is used to combine both virtual (matrix) indices $(\alpha,\beta)$ and we use the summation convention with respect to this index, a general tangent vector $\ket{\Phi[V,\{W_{\alpha}\};Q,\{R_{\alpha}\}]}$ in  $T_{\ket{\Psi[Q,\{R_{\alpha}\}]}} \mset{M}_{\mathrm{cMPS}}$ can be defined as
\begin{equation}
\begin{split}
&\ket{\Phi[V,\{W_{\alpha}\};Q,\{R_{\alpha}\}]}=\ket{\Phi^{[Q,\{R_{\alpha}\}]}[V,\{W_{\alpha}\}]}\\
&\quad=\int_{-L/2}^{+L/2}\d x\,\left(V^{i}(x) \frac{\delta\ }{\delta Q^{i}(x)}+\sum_{\beta=1}^{q}W_{\beta}^{i}(x) \frac{\delta\ }{\delta R_{\beta}^{i}(x)}\right) \ket{\Psi[Q,\{R_{\alpha}\}]}\\
&\quad=\int_{-L/2}^{+L/2}\d x\,\tr\left[B\operator{U}(-L/2,x) \left(V(x)\otimes \operator{\one}+\sum_{\beta=1}^{q}W_{\beta}(x)\otimes \hpsid_{\beta}(x)\right)\operator{U}(x,L/2)\right]\ket{\Omega}.
\end{split}\label{eq:deftangentgeneric}
\end{equation}

Because of the gauge invariance discussed in Section~\ref{s:gauge}, not all variations in $Q$ and $R_{\alpha}$ result in changes of the physical state. Consequently, not all linearly independent choices of the matrix functions $V$ and $W_{\alpha}$ result in linearly independent tangent vectors $\ket{\Phi[V,\{W_{\alpha}\};Q,\{R_{\alpha}\}]}$. Let $Q(\eta)$ and $R_{\alpha}(\eta)$ ($\forall \alpha=1,\ldots,q$) be a one-parameter family of matrix functions, so that $Q(\eta):\mset{R}\mapsto\mathbb{C}^{D\times D}:x\mapsto Q(x;\eta)$ and similarly for $R_{\alpha}(\eta)$. If we define $Q(0)=Q:x\mapsto Q(x)$, $R_{\alpha}(0)=R_{\alpha}:x\mapsto R_{\alpha}(x)$ together with $\d Q/\d \eta(0)=V:x\mapsto V(x)$ and $\d R_{\alpha}/\d \eta(0)=W_{\alpha}:x\mapsto W_{\alpha}(x)$, then we can write
\begin{equation}
\left.\frac{\d\ }{\d \eta} \ket{\Psi[Q(\eta),{R_{\alpha}(\eta)}]}\right|_{\eta=0}=\ket{\Phi[V,\{W_{\alpha}\};Q,\{R_{\alpha}\}]}.
\end{equation}
If we now choose a one-parameter family of gauge equivalent states, so that $Q(x;\eta)=g(x; \eta)^{-1}Q(x)g(x;\eta) +g(x,\eta)^{-1} \frac{\partial g(x; \eta)}{\partial x}$ and $R(x;\eta)=g(x;\eta)^{-1} R(x) g(x;\eta)$, where the one-parameter family of gauge transforms is given by $g(x;\eta)=\exp(\eta h(x))$ and $h(x)\in \algebra{gl}(\mathbb{C},D)\equiv\mathbb{C}^{D\times D}$, $\forall x\in\mset{R}$, then we can use the gauge invariance of the cMPS representation to obtain $\ket{\Psi[Q(x;\eta),R(x;\eta)]}=\ket{\Psi[Q(x),R(x)]}$ and thus
\begin{align}
\ket{\Phi[\mathscr{M}_{\Phi}^{[Q]}[h],\{\mathscr{N}_{\alpha,\Phi}^{[R_{\alpha}]}[h]\};Q,\{R_{\alpha}\}]}=0,
\end{align}
where the maps $\mathscr{M}_{\Phi}^{[Q]}$ and $\mathscr{N}_{\alpha,\Phi}^{[R_{\alpha}]}$ ($\forall \alpha=1,\ldots,N$) are given by
\begin{align}
\mathscr{M}_{\Phi}^{[Q]}[h](x)&=[Q(x),h(x)]+\frac{\d h}{\d x}(x),&\mathscr{N}^{[R_{\alpha}]}_{\alpha,\Phi}[h](x)&=[R_{\alpha}(x),h(x)].
\end{align}
The maps $\mathscr{M}_{\Phi}^{[Q]}$ and $\mathscr{N}_{\alpha,\Phi}^{[R_{\alpha}]}$ thus establish a linear homomorphism from functions $h:\mathcal{R}\to \algebra{gl}(\mathbb{C},D)\equiv\mathbb{C}^{D\times D}$ to the kernel of the representation $\ket{\Phi[V,\{W_{\alpha}\};Q,\{R_{\alpha}\}]}$ of the tangent space $\ket{\Psi[Q,\{R_{\alpha}\}]}\in  T_{\ket{\Psi[Q,\{R_{\alpha}\}]}} \mset{M}_{\mathrm{cMPS}}$. Put differently, the representation of cMPS tangent vectors has a gauge invariance under the additive transformation law $V\leftarrow V+\mathscr{M}_{\Phi}^{[Q]}[h]$ and $W_{\alpha}\leftarrow W_{\alpha}+\mathscr{N}_{\alpha,\Phi}^{[R_{\alpha}]}[h]$. In all of the above, we have considered $B$ fixed. The gauge transformation $g(x)$ then has to satisfy the boundary condition $g(+L/2) B g(-L/2)^{-1}=B$, which also imposes a boundary condition on the set of allowed functions $h(x)$, namely
\begin{equation}
h(+L/2) B - B h(-L/2) = 0.
\end{equation}
In particular, for periodic boundary conditions with $B=\one_{D}$, we obtain that the generator $h:\mset{R}\to \mathfrak{gl}(D,\mathbb{C})$ should satisfy periodic boundary conditions $h(+L/2)=h(-L/2)$.

We now restrict to the case of open boundary conditions and discard the explicit reference to the base point $\ket{\Psi[Q,\{R_{\alpha}\}]}$ in the notation of tangent vectors. To take full advantage of the gauge freedom, we noted in Section~\ref{s:gauge} that is better to include one of the boundary vectors in the set of variational parameters. We thus generalize our definition of tangent vectors by also including variations with respect to \textit{e.g.} the right boundary vector $\bm{v}_{\text{R}}$. We write
\begin{equation}
\begin{split}
&\ket{\Phi[V,\{W_{\alpha}\},\bm{w}_{\mathrm{R}}]}\\
&\qquad=\bm{w}_{\mathrm{R}}\cdot \bm{\nabla}_{\bm{v}_{\mathrm{R}}}\ket{\Psi[Q,\{R_{\alpha}\}]}\\
&\qquad\qquad+\int_{-L/2}^{+L/2}\d x\,\left(V^{i}(x) \frac{\delta\ }{\delta Q^{i}(x)}+\sum_{\beta=1}^{N}W_{\beta}^{i}(x) \frac{\delta\ }{\delta R_{\beta}^{i}(x)}\right) \ket{\Psi[Q,\{R_{\alpha}\}]}\\
&\qquad=\bm{v}_{\mathrm{L}}^\dagger \operator{U}(-L/2,+L/2) \bm{w}_{\mathrm{R}}\ket{\Omega}\\
&\qquad\qquad+\int_{-L/2}^{+L/2}\d x\,\bm{v}_{\mathrm{L}}^\dagger \operator{U}(-L/2,x) \left(V(x)\otimes \operator{\one}+\sum_{\beta=1}^{N}W_{\beta}(x)\otimes \hpsid_{\beta}(x)\right)\operator{U}(x,L/2)\bm{v}_{\mathrm{R}}\ket{\Omega}.
\end{split}\label{eq:deftangentgeneric2}
\end{equation}
Let us revisit the gauge freedom for the new tangent vectors of Eq.~\eqref{eq:deftangentgeneric2}. The state $\ket{\Phi[V,\{W_{\alpha}\},\bm{w}_{\mathrm{R}}]}$ is invariant under the additive gauge transformation $V\leftarrow V+\mathscr{M}_{\Phi}[h]$, $W_{\alpha}\leftarrow W_{\alpha}+\mathscr{N}_{\alpha,\Phi}[h]$ and $\bm{w}_{\mathrm{R}}\leftarrow \bm{w}_{\mathrm{R}} + \bm{m}_{\Phi}[h]$ with
\begin{equation}
\bm{m}_{\Phi}[h]=-h(+L/2)\bm{v}_{\mathrm{R}}.
\end{equation}
Since $\bm{v}_{\mathrm{L}}$ is still fixed, the gauge transformation has to satisfy the boundary condition $g(-L/2)=\one_{D}$, so that its generator $h(x)$ satisfies $h(-L/2)=0$.

The overlap between two tangent vectors is given by
\begin{equation}
\begin{split}
&\braket{\Phi[\overline{V},\{\overline{W}_{\alpha}\},\overline{\bm{w}_{\mathrm{R}}}]|\Phi[V',\{W'_{\alpha}\},\bm{w'}_{\mathrm{R}}]}=\bm{w}_{\mathrm{R}}^{\dagger} l(L/2) \bm{w'}_{\mathrm{R}}\\
&\qquad+\int_{-L/2}^{+L/2}\d x\, \rbraket{l(x)|\sum_{\alpha=1}^{q} W'_{\alpha}(x) \otimes \overline{W_{\alpha}(x)} | r(x)}\\
&\qquad +\int_{-L/2}^{+L/2}\d x\int_{x}^{+L/2}\d y\, \big(l(x)\big\vert\big[V'(x)\otimes 1_{D}+\sum_{\alpha=1}^{q} W'_{\alpha}(x)\otimes \overline{R_{\alpha}(x)}\big] \mathscr{P}\mathrm{e}^{\int_{x}^{y}\d z\, \voperator{T}(z)}\\
&\qquad\qquad\qquad\qquad\qquad\qquad\qquad\qquad\quad\times \big[1_{D}\otimes \overline{V(y)}+\sum_{\alpha=1}^{q}R_{\alpha}(y)\otimes \overline{W_{\alpha}(y)}\big]|r(y)\big)\\
&\qquad+\int_{-L/2}^{+L/2}\d x\int_{-L/2}^{x}\d y\, \big(l(y)\big\vert\big[1_{D}\otimes \overline{V(y)}+\sum_{\alpha=1}^{q}R_{\alpha}(y)\otimes \overline{W_{\alpha}(y)}\big] \mathscr{P}\mathrm{e}^{\int_{y}^{x}\d z\, \voperator{T}(z)}\\
&\qquad\qquad\qquad\qquad\qquad\qquad\qquad\qquad\quad\times \big[V'(x)\otimes 1_{D}+\sum_{\alpha=1}^{q}W'_{\alpha}(x)\otimes \overline{R_{\alpha}(x)}\big]\big\vert r(x)\big).
\end{split}\label{eq:phiphioverlap}
\end{equation}
It defines a metric for the manifold $\varM_{\mathrm{cMPS}}$ and features in any coordinate-invariant expression involving cMPS tangent vectors. 
We can use the gauge freedom in the representation of tangent vectors to simplify the expression above significantly. The counting argument for the gauge degrees of freedom is now less rigorous as in the discrete case. In general, we have $D^{2}$ parameters in $h(x)$ to eliminate $D^{2}$ degrees of freedom from $\{V(x),W_{1}(x),\ldots,W_{q}(x)\}$ at every point $x$. However, this is only correct if all linearly independent algebra-valued functions $h:\mset{R}\to\algebra{gl}(\mathbb{C},D)$ map to linearly independent matrix functions $[\mathscr{M}_{\Phi}^{[Q]},\{\mathscr{N}_{\alpha,\Phi}^{[R_{\alpha}]}\}]$. Let us show that by substituting $V(x)\leftarrow \tilde{V}(x)=V(x)+\mathscr{M}_{\Phi}[h](x)$ and $W_{\alpha}(x)\leftarrow \tilde{W}_{\alpha}(x)=W_{\alpha}(x)+\mathscr{N}_{\alpha,\Phi}[h](x)$ ($\forall \alpha=1,\ldots,q$), we can indeed impose $D^2$ conditions, such as the \emph{left gauge fixing condition}:
\begin{equation}
\rbra{l(x)}\left[\tilde{V}(x)\otimes \one_{D} + \sum_{n=1}^{N} \tilde{W}_{\alpha}(x)\otimes \overline{R_{\alpha}(x)}\right]=0.\label{eq:leftgaugefix}
\end{equation}
This requires that $h$ is a solution of
\begin{equation}
\frac{\d\ }{\d x}\big[l(x)h(x)\big]=\tilde{\mathscr{T}}^{(x)}\big[l(x)h(x)\big]-\left[l(x)V(x)+\sum_{\alpha=1}^{q} R_{\alpha}(x)^{\dagger} l(x) W_{\alpha}(x)\right]
\end{equation}
which together with the boundary condition $h(-L/2)=0$ results in the solution
\begin{equation}
\rbra{l(x)h(x)}=-\int_{-L/2}^{x}\d y\, \rbra{l(y)}\left[V(y)\otimes\one_{D}+\sum_{\alpha=1}^{q} W_{\alpha}(y)\otimes \overline{R}_{\alpha}(y)\right]\mathscr{P}\exp\left[\int_{y}^{x}\voperator{T}(z)\,\d z\right].
\end{equation}
This equation gives a solution for $l(x)h(x)$. We can extract $h(x)$ by multiplying with $l(x)^{-1}$ to the left. The left density matrix $l(x)$ should be positive definite and hence invertible for every $x>-L/2$. However, at $x=-L/2$ it equals $l(-L/2)=\bm{v}_{\mathrm{L}}\bm{v}_{\mathrm{L}}^{\dagger}$ and thus becomes singular. Nevertheless, the limit $\lim_{x\to-L/2} h(x)$ should be well defined since the right hand side of the equation above, which is being multiplied with $h(x)^{-1}$, will have a similar scaling.

Alternatively, we can also impose a \emph{right gauge fixing condition}
\begin{equation}
\left[V(x)\otimes \one_{D} + \sum_{\alpha=1}^{N} W_{\alpha}(x)\otimes \overline{R_{\alpha}(x)}\right]\rket{r(x)}=0.\label{eq:rightgaugefix}
\end{equation}

%Note that we can easily find a parameterization that respects these gauge fixing conditions. For the right gauge fixing condition, it is sufficient to parameterize $V$ as $V(x)=\sum_{\alpha=1}^{N} W_{\alpha}(x) r(x) R_{\alpha}(x)^{\dagger} r(x)^{-1}$ and $W_{\alpha}(x)$ can be chosen freely. Similarly, the left gauge fixing is automatically satisfied by the parameterization $V(x)=\sum_{\alpha=1}^{N} l(x)^{-1} R_{\alpha}(x)^{\dagger}  l(x) W_{\alpha}(x)$ where $W_{\alpha}(x)$ can be chosen freely. With either choice of gauge fixing conditions, only the first integral survives in Eq.~\eqref{eq:phiphioverlap} for the overlap between two tangent vectors. Hence, all non-local contributions disappear, and by further parameterizing $W_{\alpha}(x)=l(x)^{-1/2} Y(x) r(x)^{-1/2}$ and $\bm{w}_{\mathrm{R}}=l(L/2)^{-1/2} \bm{y}$, the metric is essentially reduced to the identity in terms of the variables $Y:\mset{R}\to\mathbb{C}^{D\times D}:x\mapsto Y(x)$ and $\bm{y}\in\mathbb{C}^{D}$.

Finally, we remark that the tangent space $T_{\ket{\Psi[Q,\{R_{\alpha}\}]}} \mset{M}_{\mathrm{cMPS}}$ spanned by the states of Eq.~\eqref{eq:deftangentgeneric2} contains the original cMPS $\ket{\Psi[Q,\{R_{\alpha}\}]}$, \textit{e.g.} by choosing $V=1/L$, $W_{\alpha}=0$ and $\bm{w}_{\mathrm{R}}=0$ or by choosing $V=W_{\alpha}=0$ and $\bm{w}_{\mathrm{R}}=\bm{v}_{\mathrm{R}}$. Both choices are related by a gauge transform with $h(x)=(x/L+1/2)\one_{D}$. For a general tangent vector $\ket{\Phi[V,\{W_{\alpha}\},\bm{w}_{\mathrm{R}}]}$, we obtain
\begin{equation}
\begin{split}
&\braket{\Psi[\overline{Q},\{\overline{R}_{\alpha}\}]|\Phi[V,\{W_{\alpha}\},\bm{w}_{\mathrm{R}}]}=\bm{v}_{\mathrm{R}}^{\dagger}l(L/2) \bm{w}_{\mathrm{R}}\\
&\qquad\qquad\qquad+\int_{-L/2}^{+L/2}\d x\,\rbraket{l(x)|V(x)\otimes \one_{D}+\sum_{\alpha=1}^{N} W_{\alpha}(x)\otimes \overline{R_{\alpha}(x)}|r(x)}.
\end{split}\label{eq:overlappsiphi}
\end{equation}
 If we fix the gauge according to either the left or right gauge fixing prescription, the second term cancels. We can restrict to the orthogonal complement of $\ket{\Psi[Q,\{R_{\alpha}\}]}$ in  $T_{\ket{\Psi[Q,\{R_{\alpha}\}]}} \mset{M}_{\mathrm{cMPS}}$, which is denoted as $T_{\ket{\Psi[Q,\{R_{\alpha}\}]}} \mset{M}_{\mathrm{cMPS}}^\perp$, by further imposing
\begin{equation}
\bm{v}_{\mathrm{R}}^{\dagger}l(L/2) \bm{w}_{\mathrm{R}}=0.
\end{equation}

\subsection{Uniform case}
We specialize again to the case of translation invariant systems in the thermodynamic limit. While the parameter space is now finite dimensional, it is fruitful to still consider the full tangent space to the manifold of all (translation non-invariant) cMPS at the special uniform point $\ket{\Psi(Q,\{R_{\alpha}\})}$. This boils down to allowing space-dependent matrix functions $V(x)$ and $W_{\alpha}(x)$ in the definition of the tangent vectors. We can then decompose the full tangent space into sectors $\Tplane_{\Phi_{p}}$ of momentum $p\in\mathbb{R}$ by introducing Fourier modes $V(x)=V \ec^{\ic p x}$ and $W_{\alpha}(x)=W_{\alpha}\ec^{\ic p x}$, resulting in
\begin{multline}
\ket{\Phi_{p}(V,\{W_{\alpha}\};Q,\{R_{\alpha}\})}=\ket{\Phi_{p}^{(Q,\{R_{\alpha}\})}(V,\{W_{\alpha}\})}=\\
\int_{-\infty}^{+\infty}\d x\,\ec^{\ic p x} \bm{v}_{\mathrm{L}}^{\dagger}\operator{U}(-\infty,x) \left(V\otimes \operator{\one}+\sum_{\alpha=1}^{N}W_{\alpha}\otimes \hpsid_{\alpha}(x)\right)\operator{U}(x,+\infty)\bm{v}_{\mathrm{R}}\ket{\Omega}.
\end{multline}
Note that the boundary vectors $\bm{v}_{\mathrm{L},\mathrm{R}}$ are irrelevant for the bulk properties of these states, and they are therefore not included in the set of variational parameters in the thermodynamic limit. Consequently, we also do not need to differentiate with respect to one of them in order to define the tangent space. 

We can also compute the overlap between two of these tangent vectors and obtain
\begin{displaymath}
\begin{split}
&\braket{\Phi_p(\overline{V},\{\overline{W}_{\alpha}\})|\Phi_{p'}(V',\{W'_{\alpha}\})}=\int_{-\infty}^{+\infty}\d x\, \ec^{\ic (p'-p) x}\rbraket{l|\sum_{\alpha=1}^{q} W'_{\alpha} \otimes \overline{W_{\alpha}} | r}\\
&\qquad +\int_{-\infty}^{+\infty}\d x\int_{x}^{+\infty}\d y\, \ec^{\ic (p'x - py)}\big(l\big\vert\big[V'\otimes 1_{D}+\sum_{\alpha=1}^{q} W'_{\alpha}\otimes \overline{R_{\alpha}}\big] \mathrm{e}^{(y-x)\voperator{T}}\\
&\qquad\qquad\qquad\qquad\qquad\qquad\qquad\qquad\quad\times \big[1_{D}\otimes \overline{V}+\sum_{\alpha=1}^{q}R_{\alpha}\otimes \overline{W_{\alpha}}\big]|r\big)\\
&\qquad+\int_{-\infty}^{+\infty}\d x\int_{-\infty}^{x}\d y\,\ec^{\ic(p'y-px)} \big(l\big\vert\big[1_{D}\otimes \overline{V}+\sum_{\alpha=1}^{q}R_{\alpha}\otimes \overline{W_{\alpha}}\big] \mathrm{e}^{(x-y)\voperator{T}}\\
&\qquad\qquad\qquad\qquad\qquad\qquad\qquad\qquad\quad\times \big[V'\otimes 1_{D}+\sum_{\alpha=1}^{q}W'_{\alpha}\otimes \overline{R_{\alpha}}\big]\big\vert r\big).
\end{split}
\end{displaymath}
If we again resort to the decomposition of Eq.~\eqref{eq:singulardecompositionT}, we can further evaluate this to
\begin{equation}
\begin{split}
&\braket{\Phi_p(\overline{V},\{\overline{W}_{\alpha}\})|\Phi_{p'}(V',\{W'_{\alpha}\})}=\\
&\qquad 2\pi\delta(p'-p)\Big[\rbraket{l|\sum_{\alpha=1}^{q} W'_{\alpha} \otimes \overline{W_{\alpha}} | r}\\
&\qquad\qquad\qquad +\big(l\big\vert\big[V'\otimes 1_{D}+\sum_{\alpha=1}^{q} W'_{\alpha}\otimes \overline{R_{\alpha}}\big](-\voperator{T}+\ic p)^{\mathsf{P}}\big[1_{D}\otimes \overline{V}+\sum_{\alpha=1}^{q}R_{\alpha}\otimes \overline{W_{\alpha}}\big]|r\big)\\
&\qquad\qquad\qquad +\big(l\big\vert\big[1_{D}\otimes \overline{V}+\sum_{\alpha=1}^{q}R_{\alpha}\otimes \overline{W_{\alpha}}\big] (-\voperator{T}-\ic p)^{\mathsf{P}} \big[V'\otimes 1_{D}+\sum_{\alpha=1}^{q}W'_{\alpha}\otimes \overline{R_{\alpha}}\big]\big\vert r\big)\Big]\\
&\qquad+(2\pi)^2 \delta(p) \delta(p')\big(l\big\vert\big[V'\otimes 1_{D}+\sum_{\alpha=1}^{q} W'_{\alpha}\otimes \overline{R_{\alpha}}\big]\big\vert r\big)\big(l\big\vert\big[1_{D}\otimes \overline{V}+\sum_{\alpha=1}^{q}R_{\alpha}\otimes \overline{W_{\alpha}}\big]|r\big).
\end{split}\label{eq:phipoverlap}
\end{equation}
The momentum eigenstates $\ket{\Phi_{p}(V,\{W_{\alpha}\})}$ cannot be normalized to unity in the thermodynamic limit, but rather satisfy a $\delta$-normalization. For $p=p'=0$, there is an additional divergence which is stronger than the $\delta$-normalization. It can be related to the overlap between the $\ket{\Phi_{p}(V,\{W_{\alpha}\})}$ and the original cMPS $\ket{\Psi(Q,\{R_{\alpha}\})}$, which is given by
\begin{equation}
\braket{\Psi(\overline{Q},\{\overline{R}_{\alpha}\})|\Phi_p(V,\{W_{\alpha}\})}=2\pi\delta(p) \big(l\big\vert|\big[V\otimes 1_{D}+\sum_{\alpha=1}^{q} W_{\alpha}\otimes \overline{R_{\alpha}}\big]\big\vert r\big).\label{eq:psiphipoverlap}
\end{equation}

As before, a one-parameter family of local gauge transformations $g(x;s)=\exp(sh(x))$ with $h(x)\in\algebra{gl}(D;\mathbb{C})$ induces a map to the kernel of the representation $\Phi_{p}$ of $\Tplane_{\Phi_{p}}$ by setting $h(x)=h\ec^{\ic p x}$, so that
\begin{displaymath}
\ket{\Phi_{p}(\mathscr{M}_{\Phi_{p}}^{(Q)}(h),\{\mathscr{N}_{\alpha,\Phi_{p}}^{(R_{\alpha})}(h)\};Q,\{R_{\alpha}\})}=0,
\end{displaymath}
with
\begin{align}
\mathscr{M}_{\Phi_{p}}^{(Q)}(h)&=[Q,h]+\ic p h&&\text{and}&\mathscr{N}_{\alpha,\Phi_{p}}^{(R_{\alpha})}(h)=[R_{\alpha},h].
\end{align}
We henceforth omit the superscript notation of $Q$ and $R_{\alpha}$. The dimension of the kernel of the map $\Phi_{p}$ is thus $D^{2}$-dimensional, except at $p=0$. This can easily be proven, since for every non-zero $h\in\algebra{gl}(D;\mathbb{C})$, $\mathscr{M}_{\Phi_{p}}(h)\neq 0$ or $\mathscr{N}_{\alpha,\Phi_{p}}(h)\neq 0$, $\forall \alpha=1,\ldots,N$. Indeed, suppose that $\mathscr{M}_{\Phi_{p}}(h)= 0$ and $\mathscr{N}_{\Phi_{p}}(h)=0$. Imposing that
\begin{displaymath}
\mathscr{M}_{\Phi_{p}}(h) r+\sum_{\alpha=1}^{N} \mathscr{N}_{\alpha,\Phi_{p}}(h) r R_{\alpha}^{\dagger} =0
\end{displaymath}
results in $\voperator{T} \rket{h r}=\ic p \rket{h r}$ which has no non-trivial solution except at $p=0$, where we find $h=c\one_{D}$ with $c\in\mathbb{C}$. At nonzero momenta, we can use a gauge fixing condition to reduce the number of parameters by $D^{2}$. At $p=0$, we can only reduce the number of parameters by $D^{2}-1$ through gauge fixing. But imposing orthogonality to $\ket{\Psi(Q,R)}$ manually at $p=0$ allows to discard one additional parameter. For any momentum $p$, we can uniquely fix the gauge of any tangent vector in $\Tplane_{\Phi_{p}}^{\perp}$ by setting $\rbra{l}V\otimes 1_{D} + W\otimes R=0$ or $V\otimes 1_{D} + W\otimes R\rket{r}=0$, corresponding to the left and right gauge fixing conditions respectively. It can indeed be checked that with either one of these conditions being satisfied, the overlap $\braket{\Psi(\overline{Q},\{\overline{R}_{\alpha}\})|\Phi_p(V,\{W_{\alpha}\})}$ given in Eq.~\eqref{eq:psiphipoverlap} vanishes even for $p=0$. In addition, if either gauge fixing condition is satisfied, the overlap between two tangent vectors simplifies significantly, as only the local term survives. Also note the difference with the approach for translation non-invariant systems in the previous subsection. There we could impose the left or right gauge fixing condition for any $x$, without this automatically implying that $\ket{\Phi[V,\{W_{\alpha}\},\bm{w}_{\mathrm{R}}]}\perp \ket{\Psi[Q,\{R_{\alpha}\}]}$, since a non-zero overlap between the tangent vector and the original cMPS could be encoded in the changing boundary vector $\bm{w}_{\mathrm{R}}$.

\section{Conclusion and outlook}
 This manuscript provides a detailed description of a variational class of wave functions for one-dimensional quantum field theories, that goes by the name of ``continuous matrix product states''. We reviewed different alternative constructions that produce the same class of states and have their own merits, \textit{e.g.} in offering clear hints on how to generalize this class to different settings such as open quantum systems or higher-dimensional theories. 

We illustrated how to formulate the cMPS ansatz for the most general class of theories including an arbitrary number of bosonic and fermionic particles, and were naturally led to a set of constraints that the variational parameters needed to satisfy in order to produce a finite kinetic energy density. We also discussed other physical constraints such as fermion parity. We then proceeded by explaining in detail how to compute expectation values, in particular for the case of systems with open boundary conditions. We provided some additional details for the case of systems with translation invariance, where we can use the expectation value of a correlation function to define an ultraviolet cutoff within the cMPS state.

We also discussed the important topic of gauge invariance in the cMPS representation. Finally we introduced the concept of cMPS tangent vectors, and discussed how the gauge invariance allows to represent them in such a way that the metric of the cMPS manifold simplifies tremendously. 

While we have not introduced any practical algorithms or recipes for finding cMPS approximations of ground states or for describing other physical phenomena, we have introduced all necessary definitions and concepts in order to comfortably work with cMPS. This set of definitions can now be used in follow-up papers that will focus on new algorithms. As such, the current paper provides a stepping stone that will hopefully spur more research in the context of variational methods for quantum field theories in one dimension and beyond.
   
\begin{acknowledgements}
JH acknowledges fruitful discussions with Micha\"{e}l Mari\"{e}n. This work was supported by the EU grants QUERG and QFTCMPS, by the FWF SFB grants FoQuS and ViCoM, by the DFG cluster of excellence NIM and by the cluster of excellence EXC 201 Quantum Engineering and Space-Time Research.
\end{acknowledgements}

\bibliography{paperslibrary,manuallibrary,books}

%merlin.mbs aipauth4-1.bst 2010-07-25 4.21a (PWD, AO, DPC) hacked
%Control: author (9) reversed initials
\begin{thebibliography}{41}%
\makeatletter
\providecommand \@ifxundefined [1]{%
 \@ifx{#1\undefined}
}%
\providecommand \@ifnum [1]{%
 \ifnum #1\expandafter \@firstoftwo
 \else \expandafter \@secondoftwo
 \fi
}%
\providecommand \@ifx [1]{%
 \ifx #1\expandafter \@firstoftwo
 \else \expandafter \@secondoftwo
 \fi
}%
\providecommand \natexlab [1]{#1}%
\providecommand \enquote  [1]{``#1''}%
\providecommand \bibnamefont  [1]{#1}%
\providecommand \bibfnamefont [1]{#1}%
\providecommand \citenamefont [1]{#1}%
\providecommand \href@noop [0]{\@secondoftwo}%
\providecommand \href [0]{\begingroup \@sanitize@url \@href}%
\providecommand \@href[1]{\@@startlink{#1}\@@href}%
\providecommand \@@href[1]{\endgroup#1\@@endlink}%
\providecommand \@sanitize@url [0]{\catcode `\\12\catcode `\$12\catcode
  `\&12\catcode `\#12\catcode `\^12\catcode `\_12\catcode `\%12\relax}%
\providecommand \@@startlink[1]{}%
\providecommand \@@endlink[0]{}%
\providecommand \url  [0]{\begingroup\@sanitize@url \@url }%
\providecommand \@url [1]{\endgroup\@href {#1}{\urlprefix }}%
\providecommand \urlprefix  [0]{URL }%
\providecommand \Eprint [0]{\href }%
\providecommand \doibase [0]{http://dx.doi.org/}%
\providecommand \selectlanguage [0]{\@gobble}%
\providecommand \bibinfo  [0]{\@secondoftwo}%
\providecommand \bibfield  [0]{\@secondoftwo}%
\providecommand \translation [1]{[#1]}%
\providecommand \BibitemOpen [0]{}%
\providecommand \bibitemStop [0]{}%
\providecommand \bibitemNoStop [0]{.\EOS\space}%
\providecommand \EOS [0]{\spacefactor3000\relax}%
\providecommand \BibitemShut  [1]{\csname bibitem#1\endcsname}%
\let\auto@bib@innerbib\@empty
%</preamble>
\bibitem [{\citenamefont {{Affleck}}\ \emph {et~al.}(1987)\citenamefont
  {{Affleck}}, \citenamefont {{Kennedy}}, \citenamefont {{Lieb}},\ and\
  \citenamefont {{Tasaki}}}]{1987PhRvL..59..799A}%
  \BibitemOpen
  \bibfield  {author} {\bibinfo {author} {\bibnamefont {{Affleck}},
  \bibfnamefont {I.}}, \bibinfo {author} {\bibnamefont {{Kennedy}},
  \bibfnamefont {T.}}, \bibinfo {author} {\bibnamefont {{Lieb}}, \bibfnamefont
  {E.~H.}}, \ and\ \bibinfo {author} {\bibnamefont {{Tasaki}}, \bibfnamefont
  {H.}},\ }\href@noop {} {\bibfield  {journal} {\bibinfo  {journal} {Physical
  Review Letters}\ }\textbf {\bibinfo {volume} {59}},\ \bibinfo {pages} {799}
  (\bibinfo {year} {1987})}\BibitemShut {NoStop}%
\bibitem [{\citenamefont {{Affleck}}\ \emph {et~al.}(1988)\citenamefont
  {{Affleck}}, \citenamefont {{Kennedy}}, \citenamefont {{Lieb}},\ and\
  \citenamefont {{Tasaki}}}]{1988CMaPh.115..477A}%
  \BibitemOpen
  \bibfield  {author} {\bibinfo {author} {\bibnamefont {{Affleck}},
  \bibfnamefont {I.}}, \bibinfo {author} {\bibnamefont {{Kennedy}},
  \bibfnamefont {T.}}, \bibinfo {author} {\bibnamefont {{Lieb}}, \bibfnamefont
  {E.~H.}}, \ and\ \bibinfo {author} {\bibnamefont {{Tasaki}}, \bibfnamefont
  {H.}},\ }\href@noop {} {\bibfield  {journal} {\bibinfo  {journal}
  {Communications in Mathematical Physics}\ }\textbf {\bibinfo {volume}
  {115}},\ \bibinfo {pages} {477} (\bibinfo {year} {1988})}\BibitemShut
  {NoStop}%
\bibitem [{\citenamefont {Anderson}(1967)}]{Anderson:1967aa}%
  \BibitemOpen
  \bibfield  {author} {\bibinfo {author} {\bibnamefont {Anderson},
  \bibfnamefont {P.~W.}},\ }\href {\doibase 10.1103/PhysRevLett.18.1049}
  {\bibfield  {journal} {\bibinfo  {journal} {Physical Review Letters}\
  }\textbf {\bibinfo {volume} {18}},\ \bibinfo {pages} {1049} (\bibinfo {year}
  {1967})}\BibitemShut {NoStop}%
\bibitem [{\citenamefont {{Bardeen}}, \citenamefont {{Cooper}},\ and\
  \citenamefont {{Schrieffer}}(1957)}]{1957PhRv..106..162B}%
  \BibitemOpen
  \bibfield  {author} {\bibinfo {author} {\bibnamefont {{Bardeen}},
  \bibfnamefont {J.}}, \bibinfo {author} {\bibnamefont {{Cooper}},
  \bibfnamefont {L.~N.}}, \ and\ \bibinfo {author} {\bibnamefont
  {{Schrieffer}}, \bibfnamefont {J.~R.}},\ }\href {\doibase
  10.1103/PhysRev.106.162} {\bibfield  {journal} {\bibinfo  {journal} {Physical
  Review}\ }\textbf {\bibinfo {volume} {106}},\ \bibinfo {pages} {162}
  (\bibinfo {year} {1957})}\BibitemShut {NoStop}%
\bibitem [{\citenamefont {Brockt}\ \emph {et~al.}()\citenamefont {Brockt},
  \citenamefont {Haegeman}, \citenamefont {Jennings}, \citenamefont {Osborne},\
  and\ \citenamefont {Verstraete}}]{Brockt:fk}%
  \BibitemOpen
  \bibfield  {author} {\bibinfo {author} {\bibnamefont {Brockt}, \bibfnamefont
  {C.}}, \bibinfo {author} {\bibnamefont {Haegeman}, \bibfnamefont {J.}},
  \bibinfo {author} {\bibnamefont {Jennings}, \bibfnamefont {D.}}, \bibinfo
  {author} {\bibnamefont {Osborne}, \bibfnamefont {T.~J.}}, \ and\ \bibinfo
  {author} {\bibnamefont {Verstraete}, \bibfnamefont {F.}},\ }\href@noop {}
  {\enquote {\bibinfo {title} {The continuum limit of a tensor network: A path
  integral representation},}\ }\Eprint {http://arxiv.org/abs/arXiv:1210.5401}
  {arXiv:1210.5401} \BibitemShut {NoStop}%
\bibitem [{\citenamefont {Caves}\ and\ \citenamefont
  {Milburn}(1987)}]{Caves:1987aa}%
  \BibitemOpen
  \bibfield  {author} {\bibinfo {author} {\bibnamefont {Caves}, \bibfnamefont
  {C.~M.}}\ and\ \bibinfo {author} {\bibnamefont {Milburn}, \bibfnamefont
  {G.~J.}},\ }\href {\doibase 10.1103/PhysRevA.36.5543} {\bibfield  {journal}
  {\bibinfo  {journal} {Physical Review A}\ }\textbf {\bibinfo {volume} {36}},\
  \bibinfo {pages} {5543} (\bibinfo {year} {1987})}\BibitemShut {NoStop}%
\bibitem [{\citenamefont {{Cirac}}\ and\ \citenamefont
  {{Sierra}}(2010)}]{2010PhRvB..81j4431C}%
  \BibitemOpen
  \bibfield  {author} {\bibinfo {author} {\bibnamefont {{Cirac}}, \bibfnamefont
  {J.~I.}}\ and\ \bibinfo {author} {\bibnamefont {{Sierra}}, \bibfnamefont
  {G.}},\ }\href@noop {} {\bibfield  {journal} {\bibinfo  {journal} {Physical
  Review B}\ }\textbf {\bibinfo {volume} {81}},\ \bibinfo {pages} {104431}
  (\bibinfo {year} {2010})},\ \Eprint {http://arxiv.org/abs/arXiv:0911.3029}
  {arXiv:0911.3029} \BibitemShut {NoStop}%
\bibitem [{\citenamefont {{Cirac}}\ and\ \citenamefont
  {{Verstraete}}(2009)}]{2009JPhA...42X4004C}%
  \BibitemOpen
  \bibfield  {author} {\bibinfo {author} {\bibnamefont {{Cirac}}, \bibfnamefont
  {J.~I.}}\ and\ \bibinfo {author} {\bibnamefont {{Verstraete}}, \bibfnamefont
  {F.}},\ }\href@noop {} {\bibfield  {journal} {\bibinfo  {journal} {Journal of
  Physics A Mathematical General}\ }\textbf {\bibinfo {volume} {42}},\ \bibinfo
  {pages} {4004} (\bibinfo {year} {2009})},\ \Eprint
  {http://arxiv.org/abs/arXiv:0910.1130} {arXiv:0910.1130} \BibitemShut
  {NoStop}%
\bibitem [{\citenamefont {Dubail}, \citenamefont {Read},\ and\ \citenamefont
  {Rezayi}()}]{Dubail:fk}%
  \BibitemOpen
  \bibfield  {author} {\bibinfo {author} {\bibnamefont {Dubail}, \bibfnamefont
  {J.}}, \bibinfo {author} {\bibnamefont {Read}, \bibfnamefont {N.}}, \ and\
  \bibinfo {author} {\bibnamefont {Rezayi}, \bibfnamefont {E.~H.}},\
  }\href@noop {} {\enquote {\bibinfo {title} {Edge state inner products and
  real-space entanglement spectrum of trial quantum hall states},}\ }\Eprint
  {http://arxiv.org/abs/arXiv:1207.7119} {arXiv:1207.7119} \BibitemShut
  {NoStop}%
\bibitem [{\citenamefont {{Fannes}}, \citenamefont {{Nachtergaele}},\ and\
  \citenamefont {{Werner}}(1992)}]{1992CMaPh.144..443F}%
  \BibitemOpen
  \bibfield  {author} {\bibinfo {author} {\bibnamefont {{Fannes}},
  \bibfnamefont {M.}}, \bibinfo {author} {\bibnamefont {{Nachtergaele}},
  \bibfnamefont {B.}}, \ and\ \bibinfo {author} {\bibnamefont {{Werner}},
  \bibfnamefont {R.~F.}},\ }\href@noop {} {\bibfield  {journal} {\bibinfo
  {journal} {Communications in Mathematical Physics}\ }\textbf {\bibinfo
  {volume} {144}},\ \bibinfo {pages} {443} (\bibinfo {year}
  {1992})}\BibitemShut {NoStop}%
\bibitem [{\citenamefont {Feynman}(1954)}]{Feynman:1954aa}%
  \BibitemOpen
  \bibfield  {author} {\bibinfo {author} {\bibnamefont {Feynman}, \bibfnamefont
  {R.~P.}},\ }\href {\doibase 10.1103/PhysRev.94.262} {\bibfield  {journal}
  {\bibinfo  {journal} {Physical Review}\ }\textbf {\bibinfo {volume} {94}},\
  \bibinfo {pages} {262} (\bibinfo {year} {1954})}\BibitemShut {NoStop}%
\bibitem [{\citenamefont {Feynman}\ and\ \citenamefont
  {Cohen}(1956)}]{Feynman:1956aa}%
  \BibitemOpen
  \bibfield  {author} {\bibinfo {author} {\bibnamefont {Feynman}, \bibfnamefont
  {R.~P.}}\ and\ \bibinfo {author} {\bibnamefont {Cohen}, \bibfnamefont {M.}},\
  }\href {\doibase 10.1103/PhysRev.102.1189} {\bibfield  {journal} {\bibinfo
  {journal} {Physical Review}\ }\textbf {\bibinfo {volume} {102}},\ \bibinfo
  {pages} {1189} (\bibinfo {year} {1956})}\BibitemShut {NoStop}%
\bibitem [{\citenamefont {Gross}(1961)}]{Gross:1961aa}%
  \BibitemOpen
  \bibfield  {author} {\bibinfo {author} {\bibnamefont {Gross}, \bibfnamefont
  {E.~P.}},\ }\href {\doibase 10.1007/BF02731494} {\bibfield  {journal}
  {\bibinfo  {journal} {Il Nuovo Cimento}\ }\textbf {\bibinfo {volume} {20}},\
  \bibinfo {pages} {454} (\bibinfo {year} {1961})}\BibitemShut {NoStop}%
\bibitem [{\citenamefont {{Haegeman}}\ \emph {et~al.}(2011)\citenamefont
  {{Haegeman}}, \citenamefont {{Cirac}}, \citenamefont {{Osborne}},
  \citenamefont {{Pizorn}}, \citenamefont {{Verschelde}},\ and\ \citenamefont
  {{Verstraete}}}]{2011arXiv1103.0936H}%
  \BibitemOpen
  \bibfield  {author} {\bibinfo {author} {\bibnamefont {{Haegeman}},
  \bibfnamefont {J.}}, \bibinfo {author} {\bibnamefont {{Cirac}}, \bibfnamefont
  {J.~I.}}, \bibinfo {author} {\bibnamefont {{Osborne}}, \bibfnamefont
  {T.~J.}}, \bibinfo {author} {\bibnamefont {{Pizorn}}, \bibfnamefont {I.}},
  \bibinfo {author} {\bibnamefont {{Verschelde}}, \bibfnamefont {H.}}, \ and\
  \bibinfo {author} {\bibnamefont {{Verstraete}}, \bibfnamefont {F.}},\
  }\href@noop {} {\bibfield  {journal} {\bibinfo  {journal} {Physical Review
  Letters}\ }\textbf {\bibinfo {volume} {107}},\ \bibinfo {pages} {070601}
  (\bibinfo {year} {2011})},\ \Eprint {http://arxiv.org/abs/arXiv:1103.0936}
  {arXiv:1103.0936} \BibitemShut {NoStop}%
\bibitem [{\citenamefont {Haegeman}\ \emph {et~al.}()\citenamefont {Haegeman},
  \citenamefont {Mari\"{e}n}, \citenamefont {Osborne},\ and\ \citenamefont
  {Verstraete}}]{Haegeman:fk}%
  \BibitemOpen
  \bibfield  {author} {\bibinfo {author} {\bibnamefont {Haegeman},
  \bibfnamefont {J.}}, \bibinfo {author} {\bibnamefont {Mari\"{e}n},
  \bibfnamefont {M.}}, \bibinfo {author} {\bibnamefont {Osborne}, \bibfnamefont
  {T.~J.}}, \ and\ \bibinfo {author} {\bibnamefont {Verstraete}, \bibfnamefont
  {F.}},\ }\href@noop {} {\ }\Eprint {http://arxiv.org/abs/arXiv:1210.7710}
  {arXiv:1210.7710} \BibitemShut {NoStop}%
\bibitem [{\citenamefont {{Haegeman}}\ \emph {et~al.}(2012)\citenamefont
  {{Haegeman}}, \citenamefont {{Pirvu}}, \citenamefont {{Weir}}, \citenamefont
  {{Cirac}}, \citenamefont {{Osborne}}, \citenamefont {{Verschelde}},\ and\
  \citenamefont {{Verstraete}}}]{2011arXiv1103.2286H}%
  \BibitemOpen
  \bibfield  {author} {\bibinfo {author} {\bibnamefont {{Haegeman}},
  \bibfnamefont {J.}}, \bibinfo {author} {\bibnamefont {{Pirvu}}, \bibfnamefont
  {B.}}, \bibinfo {author} {\bibnamefont {{Weir}}, \bibfnamefont {D.~J.}},
  \bibinfo {author} {\bibnamefont {{Cirac}}, \bibfnamefont {J.~I.}}, \bibinfo
  {author} {\bibnamefont {{Osborne}}, \bibfnamefont {T.~J.}}, \bibinfo {author}
  {\bibnamefont {{Verschelde}}, \bibfnamefont {H.}}, \ and\ \bibinfo {author}
  {\bibnamefont {{Verstraete}}, \bibfnamefont {F.}},\ }\href@noop {} {\bibfield
   {journal} {\bibinfo  {journal} {Physical Review B}\ }\textbf {\bibinfo
  {volume} {85}},\ \bibinfo {pages} {100408(R)} (\bibinfo {year} {2012})},\
  \Eprint {http://arxiv.org/abs/arXiv:1103.2286} {arXiv:1103.2286} \BibitemShut
  {NoStop}%
\bibitem [{\citenamefont {Haldane}(1983{\natexlab{a}})}]{Haldane:1983aa}%
  \BibitemOpen
  \bibfield  {author} {\bibinfo {author} {\bibnamefont {Haldane}, \bibfnamefont
  {F.~D.~M.}},\ }\href {\doibase 10.1016/0375-9601(83)90631-X} {\bibfield
  {journal} {\bibinfo  {journal} {Physics Letters A}\ }\textbf {\bibinfo
  {volume} {93}},\ \bibinfo {pages} {464} (\bibinfo {year}
  {1983}{\natexlab{a}})}\BibitemShut {NoStop}%
\bibitem [{\citenamefont {Haldane}(1983{\natexlab{b}})}]{Haldane:1983ab}%
  \BibitemOpen
  \bibfield  {author} {\bibinfo {author} {\bibnamefont {Haldane}, \bibfnamefont
  {F.~D.~M.}},\ }\href {\doibase 10.1103/PhysRevLett.50.1153} {\bibfield
  {journal} {\bibinfo  {journal} {Physical Review Letters}\ }\textbf {\bibinfo
  {volume} {50}},\ \bibinfo {pages} {1153} (\bibinfo {year}
  {1983}{\natexlab{b}})}\BibitemShut {NoStop}%
\bibitem [{\citenamefont {Laughlin}(1983)}]{PhysRevLett.50.1395}%
  \BibitemOpen
  \bibfield  {author} {\bibinfo {author} {\bibnamefont {Laughlin},
  \bibfnamefont {R.~B.}},\ }\href {\doibase 10.1103/PhysRevLett.50.1395}
  {\bibfield  {journal} {\bibinfo  {journal} {Phys. Rev. Lett.}\ }\textbf
  {\bibinfo {volume} {50}},\ \bibinfo {pages} {1395} (\bibinfo {year}
  {1983})}\BibitemShut {NoStop}%
\bibitem [{\citenamefont {{Lindblad}}(1976)}]{1976CMaPh..48..119L}%
  \BibitemOpen
  \bibfield  {author} {\bibinfo {author} {\bibnamefont {{Lindblad}},
  \bibfnamefont {G.}},\ }\href {\doibase 10.1007/BF01608499} {\bibfield
  {journal} {\bibinfo  {journal} {Communications in Mathematical Physics}\
  }\textbf {\bibinfo {volume} {48}},\ \bibinfo {pages} {119} (\bibinfo {year}
  {1976})}\BibitemShut {NoStop}%
\bibitem [{\citenamefont {{Milsted}}\ \emph {et~al.}(2012)\citenamefont
  {{Milsted}}, \citenamefont {{Haegeman}}, \citenamefont {{Osborne}},\ and\
  \citenamefont {{Verstraete}}}]{2012arXiv1207.0691M}%
  \BibitemOpen
  \bibfield  {author} {\bibinfo {author} {\bibnamefont {{Milsted}},
  \bibfnamefont {A.}}, \bibinfo {author} {\bibnamefont {{Haegeman}},
  \bibfnamefont {J.}}, \bibinfo {author} {\bibnamefont {{Osborne}},
  \bibfnamefont {T.~J.}}, \ and\ \bibinfo {author} {\bibnamefont
  {{Verstraete}}, \bibfnamefont {F.}},\ }\href@noop {} {\  (\bibinfo {year}
  {2012})},\ \Eprint {http://arxiv.org/abs/arXiv:1207.0691} {arXiv:1207.0691}
  \BibitemShut {NoStop}%
\bibitem [{\citenamefont {Moore}\ and\ \citenamefont
  {Read}(1991)}]{Moore1991362}%
  \BibitemOpen
  \bibfield  {author} {\bibinfo {author} {\bibnamefont {Moore}, \bibfnamefont
  {G.}}\ and\ \bibinfo {author} {\bibnamefont {Read}, \bibfnamefont {N.}},\
  }\href {\doibase 10.1016/0550-3213(91)90407-O} {\bibfield  {journal}
  {\bibinfo  {journal} {Nuclear Physics B}\ }\textbf {\bibinfo {volume}
  {360}},\ \bibinfo {pages} {362} (\bibinfo {year} {1991})}\BibitemShut
  {NoStop}%
\bibitem [{\citenamefont {{Nielsen}}, \citenamefont {{Sierra}},\ and\
  \citenamefont {{Cirac}}(2011)}]{2011PhRvA..83e3807N}%
  \BibitemOpen
  \bibfield  {author} {\bibinfo {author} {\bibnamefont {{Nielsen}},
  \bibfnamefont {A.~E.~B.}}, \bibinfo {author} {\bibnamefont {{Sierra}},
  \bibfnamefont {G.}}, \ and\ \bibinfo {author} {\bibnamefont {{Cirac}},
  \bibfnamefont {J.~I.}},\ }\href@noop {} {\bibfield  {journal} {\bibinfo
  {journal} {Physical Review A}\ }\textbf {\bibinfo {volume} {83}},\ \bibinfo
  {pages} {053807} (\bibinfo {year} {2011})},\ \Eprint
  {http://arxiv.org/abs/arXiv:1103.2205} {arXiv:1103.2205} \BibitemShut
  {NoStop}%
\bibitem [{Note1()}]{Note1}%
  \BibitemOpen
  \bibinfo {note} {CMPS still obey the infrared orthogonality catastrophe when
  formulated in the thermodynamic limit (see Section~\ref {s:ti})}\BibitemShut
  {NoStop}%
\bibitem [{Note2()}]{Note2}%
  \BibitemOpen
  \bibinfo {note} {If there is no insertion at the same position, we can always
  insert a unit operator $\protect \openone _D$}\BibitemShut {NoStop}%
\bibitem [{Note3()}]{Note3}%
  \BibitemOpen
  \bibinfo {note} {While we mentioned in Section~\ref {s:bc} that we always
  assume the matrix functions $Q$ and $R_{\alpha }$ to satisfy the proper
  boundary conditions, we do not have to use the condition in Eq.~\protect
  \textup {\hbox {\mathsurround \z@ \protect \normalfont (\ignorespaces \ref
  {eq:qropenbc}\unskip \@@italiccorr )}} at any point in deriving the
  expectation value of the Hamiltonian $\protect \ensuremath {\protect
  \mathaccentV {hat}05E{H}}$ in Eq.~\protect \textup {\hbox {\mathsurround \z@
  \protect \normalfont (\ignorespaces \ref {eq:generichamiltonian}\unskip
  \@@italiccorr )}}.}\BibitemShut {Stop}%
\bibitem [{Note4()}]{Note4}%
  \BibitemOpen
  \bibinfo {note} {While we take a standard matrix logarithm, it also makes
  sense to define the linear maps $\protect \EuScript {T}$, $\protect
  \mathaccentV {tilde}07E{\protect \EuScript {T}}$ as the logarithm of ---or
  the generator for--- the completely positive maps $\protect \EuScript {E}$
  and $\protect \mathaccentV {tilde}07E{\protect \EuScript {E}}$ associated to
  the left or right action of $\protect \ensuremath {\protect \mathbb {E}}$.
  However, not all completely positive maps have a natural logarithm associated
  to it, as was shown in Ref.~\protect \rev@citealp
  {2008CMaPh.279..147W}.}\BibitemShut {Stop}%
\bibitem [{\citenamefont {{Osborne}}, \citenamefont {{Eisert}},\ and\
  \citenamefont {{Verstraete}}(2010)}]{2010PhRvL.105z0401O}%
  \BibitemOpen
  \bibfield  {author} {\bibinfo {author} {\bibnamefont {{Osborne}},
  \bibfnamefont {T.~J.}}, \bibinfo {author} {\bibnamefont {{Eisert}},
  \bibfnamefont {J.}}, \ and\ \bibinfo {author} {\bibnamefont {{Verstraete}},
  \bibfnamefont {F.}},\ }\href@noop {} {\bibfield  {journal} {\bibinfo
  {journal} {Physical Review Letters}\ }\textbf {\bibinfo {volume} {105}},\
  \bibinfo {pages} {260401} (\bibinfo {year} {2010})},\ \Eprint
  {http://arxiv.org/abs/arXiv:1005.1268} {arXiv:1005.1268} \BibitemShut
  {NoStop}%
\bibitem [{\citenamefont {{{\"O}stlund}}\ and\ \citenamefont
  {{Rommer}}(1995)}]{1995PhRvL..75.3537O}%
  \BibitemOpen
  \bibfield  {author} {\bibinfo {author} {\bibnamefont {{{\"O}stlund}},
  \bibfnamefont {S.}}\ and\ \bibinfo {author} {\bibnamefont {{Rommer}},
  \bibfnamefont {S.}},\ }\href@noop {} {\bibfield  {journal} {\bibinfo
  {journal} {Physical Review Letters}\ }\textbf {\bibinfo {volume} {75}},\
  \bibinfo {pages} {3537} (\bibinfo {year} {1995})},\ \Eprint
  {http://arxiv.org/abs/arXiv:cond-mat/9503107} {arXiv:cond-mat/9503107}
  \BibitemShut {NoStop}%
\bibitem [{\citenamefont {{Perez-Garcia}}\ \emph {et~al.}(2007)\citenamefont
  {{Perez-Garcia}}, \citenamefont {{Verstraete}}, \citenamefont {{Wolf}},\ and\
  \citenamefont {{Cirac}}}]{2006quant.ph..8197P}%
  \BibitemOpen
  \bibfield  {author} {\bibinfo {author} {\bibnamefont {{Perez-Garcia}},
  \bibfnamefont {D.}}, \bibinfo {author} {\bibnamefont {{Verstraete}},
  \bibfnamefont {F.}}, \bibinfo {author} {\bibnamefont {{Wolf}}, \bibfnamefont
  {M.~M.}}, \ and\ \bibinfo {author} {\bibnamefont {{Cirac}}, \bibfnamefont
  {J.~I.}},\ }\href@noop {} {\bibfield  {journal} {\bibinfo  {journal} {Quantum
  Information and Computation}\ }\textbf {\bibinfo {volume} {7}},\ \bibinfo
  {pages} {401} (\bibinfo {year} {2007})},\ \Eprint
  {http://arxiv.org/abs/arXiv:quant-ph/0608197} {arXiv:quant-ph/0608197}
  \BibitemShut {NoStop}%
\bibitem [{\citenamefont {{Pirvu}}, \citenamefont {{Haegeman}},\ and\
  \citenamefont {{Verstraete}}(2012)}]{2012PhRvB..85c5130P}%
  \BibitemOpen
  \bibfield  {author} {\bibinfo {author} {\bibnamefont {{Pirvu}}, \bibfnamefont
  {B.}}, \bibinfo {author} {\bibnamefont {{Haegeman}}, \bibfnamefont {J.}}, \
  and\ \bibinfo {author} {\bibnamefont {{Verstraete}}, \bibfnamefont {F.}},\
  }\href {\doibase 10.1103/PhysRevB.85.035130} {\bibfield  {journal} {\bibinfo
  {journal} {Physical Review B}\ }\textbf {\bibinfo {volume} {85}},\ \bibinfo
  {pages} {035130} (\bibinfo {year} {2012})},\ \Eprint
  {http://arxiv.org/abs/arXiv:1103.2735} {arXiv:1103.2735} \BibitemShut
  {NoStop}%
\bibitem [{\citenamefont {Pitaevskii}(1961)}]{Pitaevskii:1961aa}%
  \BibitemOpen
  \bibfield  {author} {\bibinfo {author} {\bibnamefont {Pitaevskii},
  \bibfnamefont {L.~P.}},\ }\href@noop {} {\bibfield  {journal} {\bibinfo
  {journal} {Soviet Journal of Experimental and Theoretical Physics}\ }\textbf
  {\bibinfo {volume} {13}},\ \bibinfo {pages} {451} (\bibinfo {year}
  {1961})}\BibitemShut {NoStop}%
\bibitem [{\citenamefont {{Pollmann}}\ \emph {et~al.}(2010)\citenamefont
  {{Pollmann}}, \citenamefont {{Turner}}, \citenamefont {{Berg}},\ and\
  \citenamefont {{Oshikawa}}}]{2010PhRvB..81f4439P}%
  \BibitemOpen
  \bibfield  {author} {\bibinfo {author} {\bibnamefont {{Pollmann}},
  \bibfnamefont {F.}}, \bibinfo {author} {\bibnamefont {{Turner}},
  \bibfnamefont {A.~M.}}, \bibinfo {author} {\bibnamefont {{Berg}},
  \bibfnamefont {E.}}, \ and\ \bibinfo {author} {\bibnamefont {{Oshikawa}},
  \bibfnamefont {M.}},\ }\href {\doibase 10.1103/PhysRevB.81.064439} {\bibfield
   {journal} {\bibinfo  {journal} {Physical Review B}\ }\textbf {\bibinfo
  {volume} {81}},\ \bibinfo {pages} {064439} (\bibinfo {year} {2010})},\
  \Eprint {http://arxiv.org/abs/arXiv:0910.1811} {arXiv:0910.1811} \BibitemShut
  {NoStop}%
\bibitem [{\citenamefont {{Rommer}}\ and\ \citenamefont
  {{{\"O}stlund}}(1997)}]{1997PhRvB..55.2164R}%
  \BibitemOpen
  \bibfield  {author} {\bibinfo {author} {\bibnamefont {{Rommer}},
  \bibfnamefont {S.}}\ and\ \bibinfo {author} {\bibnamefont {{{\"O}stlund}},
  \bibfnamefont {S.}},\ }\href@noop {} {\bibfield  {journal} {\bibinfo
  {journal} {Physical Review B}\ }\textbf {\bibinfo {volume} {55}},\ \bibinfo
  {pages} {2164} (\bibinfo {year} {1997})},\ \Eprint
  {http://arxiv.org/abs/arXiv:cond-mat/9606213} {arXiv:cond-mat/9606213}
  \BibitemShut {NoStop}%
\bibitem [{\citenamefont {{Sch{\"o}n}}\ \emph {et~al.}(2005)\citenamefont
  {{Sch{\"o}n}}, \citenamefont {{Solano}}, \citenamefont {{Verstraete}},
  \citenamefont {{Cirac}},\ and\ \citenamefont {{Wolf}}}]{2005PhRvL..95k0503S}%
  \BibitemOpen
  \bibfield  {author} {\bibinfo {author} {\bibnamefont {{Sch{\"o}n}},
  \bibfnamefont {C.}}, \bibinfo {author} {\bibnamefont {{Solano}},
  \bibfnamefont {E.}}, \bibinfo {author} {\bibnamefont {{Verstraete}},
  \bibfnamefont {F.}}, \bibinfo {author} {\bibnamefont {{Cirac}}, \bibfnamefont
  {J.~I.}}, \ and\ \bibinfo {author} {\bibnamefont {{Wolf}}, \bibfnamefont
  {M.~M.}},\ }\href@noop {} {\bibfield  {journal} {\bibinfo  {journal}
  {Physical Review Letters}\ }\textbf {\bibinfo {volume} {95}},\ \bibinfo
  {pages} {110503} (\bibinfo {year} {2005})},\ \Eprint
  {http://arxiv.org/abs/arXiv:quant-ph/0501096} {arXiv:quant-ph/0501096}
  \BibitemShut {NoStop}%
\bibitem [{\citenamefont {{Verstraete}}()}]{qgp}%
  \BibitemOpen
  \bibfield  {author} {\bibinfo {author} {\bibnamefont {{Verstraete}},
  \bibfnamefont {F.}},\ }\href@noop {} {}\bibinfo {note} {In
  preparation}\BibitemShut {NoStop}%
\bibitem [{\citenamefont {{Verstraete}}\ and\ \citenamefont
  {{Cirac}}(2010)}]{2010PhRvL.104s0405V}%
  \BibitemOpen
  \bibfield  {author} {\bibinfo {author} {\bibnamefont {{Verstraete}},
  \bibfnamefont {F.}}\ and\ \bibinfo {author} {\bibnamefont {{Cirac}},
  \bibfnamefont {J.~I.}},\ }\href@noop {} {\bibfield  {journal} {\bibinfo
  {journal} {Physical Review Letters}\ }\textbf {\bibinfo {volume} {104}},\
  \bibinfo {pages} {190405} (\bibinfo {year} {2010})},\ \Eprint
  {http://arxiv.org/abs/arXiv:1002.1824} {arXiv:1002.1824} \BibitemShut
  {NoStop}%
\bibitem [{\citenamefont {{Verstraete}}, \citenamefont {{Murg}},\ and\
  \citenamefont {{Cirac}}(2008)}]{2008AdPhy..57..143V}%
  \BibitemOpen
  \bibfield  {author} {\bibinfo {author} {\bibnamefont {{Verstraete}},
  \bibfnamefont {F.}}, \bibinfo {author} {\bibnamefont {{Murg}}, \bibfnamefont
  {V.}}, \ and\ \bibinfo {author} {\bibnamefont {{Cirac}}, \bibfnamefont
  {J.~I.}},\ }\href@noop {} {\bibfield  {journal} {\bibinfo  {journal}
  {Advances in Physics}\ }\textbf {\bibinfo {volume} {57}},\ \bibinfo {pages}
  {143} (\bibinfo {year} {2008})},\ \Eprint
  {http://arxiv.org/abs/arXiv:0907.2796} {arXiv:0907.2796} \BibitemShut
  {NoStop}%
\bibitem [{\citenamefont {{White}}(1992)}]{1992PhRvL..69.2863W}%
  \BibitemOpen
  \bibfield  {author} {\bibinfo {author} {\bibnamefont {{White}}, \bibfnamefont
  {S.~R.}},\ }\href@noop {} {\bibfield  {journal} {\bibinfo  {journal}
  {Physical Review Letters}\ }\textbf {\bibinfo {volume} {69}},\ \bibinfo
  {pages} {2863} (\bibinfo {year} {1992})}\BibitemShut {NoStop}%
\bibitem [{\citenamefont {{White}}(1993)}]{1993PhRvB..4810345W}%
  \BibitemOpen
  \bibfield  {author} {\bibinfo {author} {\bibnamefont {{White}}, \bibfnamefont
  {S.~R.}},\ }\href@noop {} {\bibfield  {journal} {\bibinfo  {journal}
  {Physical Review B}\ }\textbf {\bibinfo {volume} {48}},\ \bibinfo {pages}
  {10345} (\bibinfo {year} {1993})}\BibitemShut {NoStop}%
\bibitem [{\citenamefont {{Wolf}}\ and\ \citenamefont
  {{Cirac}}(2008)}]{2008CMaPh.279..147W}%
  \BibitemOpen
  \bibfield  {author} {\bibinfo {author} {\bibnamefont {{Wolf}}, \bibfnamefont
  {M.~M.}}\ and\ \bibinfo {author} {\bibnamefont {{Cirac}}, \bibfnamefont
  {J.~I.}},\ }\href {\doibase 10.1007/s00220-008-0411-y} {\bibfield  {journal}
  {\bibinfo  {journal} {Communications in Mathematical Physics}\ }\textbf
  {\bibinfo {volume} {279}},\ \bibinfo {pages} {147} (\bibinfo {year}
  {2008})},\ \Eprint {http://arxiv.org/abs/arXiv:math-ph/0611057}
  {arXiv:math-ph/0611057} \BibitemShut {NoStop}%
\end{thebibliography}%
\bibliographystyle{aipauth4-1}
\hphantom{\cite{2008CMaPh.279..147W}.}

\appendix
\section{A useful formula}
\label{a:formula}
Consider an operator $\operator{U}(x,y)$ defined as
\begin{equation}
\operator{U}(x,y)=\Pexp\left[\int_x^y \operator{A}(z)\,\d z\right],
\end{equation}
where $\operator{A}$ is not necessarily antihermitian. 
This operator satisfies
\begin{align}
\frac{\d\ }{\d x} \operator{U}(x,y)&=-\operator{A}(x) \operator{U}(x,y),&
\frac{\d\ }{\d y} \operator{U}(x,y)&=+\operator{U}(x,y) \operator{A}(y).\label{eq:diffU}
\end{align}
For the derivatives of the inverse operator $\operator{U}(x,y)^{-1}$ we can use the general result
\begin{align}
\frac{\d\ }{\d x} \operator{U}(x,y)^{-1} &= - \operator{U}(x,y)^{-1} \left(\frac{\d\ }{\d x} \operator{U}(x,y)\right) \operator{U}(x,y)^{-1}=+\operator{U}(x,y)^{-1} \operator{A}(x),\\
\frac{\d\ }{\d y} \operator{U}(x,y)^{-1} &= - \operator{U}(x,y)^{-1} \left(\frac{\d\ }{\d y} \operator{U}(x,y)\right) \operator{U}(x,y)^{-1}=- \operator{A}(y)\operator{U}(x,y)^{-1},\\
\end{align}

Now define the following operator quantity depending on an arbitrary operator $\operator{B}$
\begin{equation}
\operator{C}(x,y)=\operator{U}(x,y)\operator{B} \operator{U}(x,y)^{-1}.
\end{equation}
By taking the derivative with respect to $y$, we obtain
\begin{displaymath}
\frac{\d\ }{\d y}\operator{C}(x,y)=\operator{U}(x,y)\left[\operator{A}(y),\operator{B}\right] \operator{U}(x,y)^{-1}.
\end{displaymath}
Integrating $\d \operator{C}(x,z) /\d z$ for $z$ from $x$ to $y$ and making use of the initial value $\operator{C}(x,x)=\operator{B}$ results in
\begin{equation}
\operator{C}(x,y)=\operator{B}+\int_x^y \operator{U}(x,z) \left[\operator{A}(z),\operator{B}\right]\operator{U}(x,z)^{-1}\, \d z.
\end{equation}
We then multiply this equality with $\operator{U}(x,y)$ to the right and make use of the obvious identity $\operator{U}(x,y)=\operator{U}(x,z) \operator{U}(z,y)$ for any $x<z<y$ in the integral of the right hand side in order to obtain our final result
\begin{equation}
\left[\operator{U}(x,y),\operator{B}\right]=\int_x^y \operator{U}(x,z) \left[\operator{A}(z),\operator{B}\right]\operator{U}(z,y)\,\d z.\label{eq:commutatorequality}
\end{equation}

We can further generalize this result. Suppose we have two operators $\operator{U}_{\pm}(x,y)$ defined as 
\begin{equation}
\operator{U}_{\pm}(x,y)=\Pexp\left[\int_x^y \left\{\operator{A}_1(z) \pm \operator{A}_2(z)\right\}\,\d z\right],
\end{equation}
for arbitrary $\operator{A}_{1,2}(z)$. If we consider the quantity
\begin{equation}
\operator{C}(x,y)=\operator{U}_{-}(x,y)\operator{B} \operator{U}_{+}(x,y)^{-1},
\end{equation}
then we obtain
\begin{displaymath}
\frac{\d\ }{\d y}\operator{C}(x,y)=\operator{U}_{-}(x,y)\left(\left[\operator{A}_1(y),\operator{B}\right]-\left\{\operator{A}_{2}(y),\operator{B}\right\}\right) \operator{U}(x,y)_{+}^{-1},
\end{displaymath}
using a similar derivation. Continuing along the same line results in
\begin{equation}
\operator{B}\operator{U}_{+}(x,y) -\operator{U}_{-}(x,y)\operator{B} = \int_x^y \operator{U}_{-}(x,x)\left(\left[\operator{B},\operator{A}_1(z)\right]+\left\{\operator{B},\operator{A}_2(z)\right\}\right)\operator{U}_{+}(z,y)\,\d z.\label{eq:commutatorequalitygeneralized}
\end{equation}

\section{Higher order regularity conditions}
\label{a:higherorderregularity}
In this appendix we derive additional regularity conditions by considering higher derivatives of the field operators acting on the ground state. Throughout this appendix, we assume that Eq.~\eqref{eq:regcondition} is fulfilled and $R_{\alpha}(x)$ has well-behaved higher order derivatives.
We now consider the state $(\d^{2}\hpsi_{\alpha}(x)/ \d x^{2}) \ket{\Psi[Q,\{R_{\beta}\}]}$, which contains a contribution with infinite norm unless
\begin{equation}
\left[\frac{\d R_\alpha}{\d x}(x) +[Q(x),R_\alpha(x)], R_{\beta}(x)\right]_{\mp}=0,\label{eq:regcondition2}
\end{equation}
where $[\cdot,\cdot]_{\mp}$ is a commutator ($-$) or anticommutator ($+$) for $\eta_{\alpha,\beta}=\pm 1$. If $Q$ and $R_{\alpha}$ obey all equations to have a `well defined' derivative up to order $n$, so that the state $(\d^{n}\hpsi(x)/\d x^{n})\ket{\Psi[Q,\{R_{\beta}\}]}$ is normalizable, the sufficient condition to eliminate all harmful contributions from $(\d^{n+1}\hpsi(x)/\d x^{n+1})\ket{\Psi[Q,\{R_{\beta}\}]}$ is
\begin{multline}
\bigg[\frac{\d^{n}\ }{\d x^{n}}R_\alpha(x) +\frac{\mathrm{d}^{n-1}\ }{\mathrm{d} x^{n-1}}[Q(x),R_\alpha(x)]+\frac{\mathrm{d}^{n-2}\ }{\mathrm{d} x^{n-2}}[Q(x),[Q(x),R_\alpha(x)]]\\
 + \ldots + [Q(x),[\ldots,[Q(x),R(x)]] \ldots ] , R_{\beta}(x)\bigg]_{\mp}=0.\label{eq:regconditionn}
\end{multline}

We can also impose regularity of the mixed derivatives of the $N$-particle wave function, by first evaluating $\hpsi_{\alpha}(x)\hpsi_{\beta}(y)\ket{\Psi[Q,\{R_{\gamma}\}]}$
\begin{multline*}
\hpsi_{\alpha}(x)\hpsi_{\beta}(y)\ket{\Psi[Q,\{R_{\gamma}\}]}=\\
\theta(y-x)\tr\left[B \operator{U}_{\alpha,\beta}(-L/2,x) \eta_{\beta,\alpha}R_{\alpha}(x) \operator{U}_{\beta}(x,y)R_{\beta}(y)\operator{U}(y,+L/2)\right]\ket{\Omega}\\
+\theta(x-y)\tr\left[B \operator{U}_{\alpha,\beta}(-L/2,y) R_{\beta}(y) \operator{U}_{\alpha}(y,x)R_{\alpha}(x)\operator{U}(x,+L/2)\right]\ket{\Omega}
\end{multline*}
where a new set of operators $\operator{U}_{\alpha,\beta}(x,y)$ ($\alpha,\beta=1,\ldots,q$) was introduced as
\begin{equation}
\operator{U}_{\alpha,\beta}(x,y)=\mathscr{P} \exp\left[\int_{x}^{y}\d z\, \left\{Q(z)\otimes \operator{\one} + \sum_{\gamma=1}^{q}\eta_{\alpha,\gamma}\eta_{\beta,\gamma}R_{\gamma}(z)\otimes \hpsid_{\gamma}(z)\right\}\right]\label{eq:defUalphabeta}.
\end{equation}
Note that the regularity condition in Eq.~\eqref{eq:regcondition} is sufficient for the annihilation of two particles $\hpsi_{\alpha}(x)\hpsi_{\beta}(y)\ket{\Psi[Q,\{R_{\gamma}\}]}$ to be continuous at $x=y$. By first differentiating with respect to $x$, we obtain
\begin{multline*}
\left(\frac{\d \hpsi_{\alpha}}{\d x}(x)\right)\hpsi_{\beta}(y)\ket{\Psi[Q,\{R_{\gamma}\}]}\\
\shoveleft{\quad=\theta(y-x)\tr\Bigg[B \operator{U}_{\alpha,\beta}(-L/2,x) \eta_{\beta,\alpha}\bigg(\frac{\d R_{\alpha}}{\d x}(x) +\big[Q(x),R_{\alpha}(x)\big]\bigg)}\\
\shoveright{\times\operator{U}_{\beta}(x,y)R_{\beta}(y)\operator{U}(y,+L/2)\Bigg]\ket{\Omega}\ \ }\\
\shoveleft{\quad\quad+\theta(x-y)\tr\Bigg[B \operator{U}_{\alpha,\beta}(-L/2,y) R_{\beta}(y) \operator{U}_{\alpha}(y,x)}\\
\times\bigg(\frac{\d R_{\alpha}}{\d x}(x) +[Q(x),R_{\alpha}(x)]\bigg)\operator{U}(x,+L/2)\Bigg]\ket{\Omega},
\end{multline*}
where we have assumed the regularity condition in Eq.~\eqref{eq:regcondition} to hold. This allows one to eliminate the fixed insertion of particles at position $x$ as well as the terms obtained from differentiating the Heaviside functions (\textit{i.e.}\ the terms proportional to $\delta(x-y)$). Such terms would indeed arise if $\hpsi_{\alpha}(x)\hpsi_{\beta}(y)\ket{\Psi[Q,\{R_{\gamma}\}]}$ were not continuous at $x=y$. If we now also differentiate with respect to $y$, we obtain a divergent contribution
\begin{displaymath}
-\delta(x-y)\tr\left[B \operator{W}_{\alpha,\beta}(-L/2,x) \left[R_{\beta}(x),\frac{\d R_{\alpha}}{\d x}(x) +[Q(x),R_{\alpha}(x)]\right]_{\mp}\operator{U}(x,+L/2)\right]\ket{\Omega}.
\end{displaymath}
If we differentiated with respect to $y$ first, and then to $x$, the divergent contribution is
\begin{displaymath}
\delta(x-y)\tr\left[B \operator{W}_{\alpha,\beta}(-L/2,x) \left[\frac{\d R_{\beta}}{\d x}(x) +[Q(x),R_{\beta}(x)],R_{\alpha}(x)\right]_{\mp}\operator{U}(x,+L/2)\right]\ket{\Omega}.
\end{displaymath}
Since we are working under assumption of the regularity condition $[R_{\beta}(x),R_{\alpha}(x)]_{\mp}=0$ [Eq.~\eqref{eq:regcondition}], it is easy to show that $[R_{\beta}(x),\d R_{\alpha}(x)/\d x]_{\mp}=-[\d R_{\beta}(x)/\d x,R_{\alpha}(x)]_{\mp}$ and also $[R_{\beta}(x),[Q(x),R_{\alpha}(x)]]_{\mp}=-[[Q(x),R_{\beta}(x)],R_{\alpha}(x)]_{\mp}$, so that both diverging contributions are equal. By imposing
\begin{equation}
\left[\frac{\d R_{\beta}}{\d x}(x) +[Q(x),R_{\beta}(x)],R_{\alpha}(x)\right]_{\mp}=-\left[R_{\beta}(x),\frac{\d R_{\alpha}}{\d x}(x) +[Q(x),R_{\alpha}(x)]\right]_{\mp}=0\label{eq:regmixed}
\end{equation}
the mixed derivative $(\d \hpsi_{\alpha}(x)/\d x)(\d \hpsi_{\beta}(y)/\d y)\ket{\Psi[Q(x),\{R_{\gamma}\}]}$ is well defined and normalizable. Note that Eq.~\eqref{eq:regmixed} is identical to Eq.~\eqref{eq:regcondition2}, so that regularity of the mixed product of two first order derivatives is guaranteed if the second order derivative is regular, or vice versa.

The higher order regularity conditions derived in this appendix put very strong constraints on $Q$ and $R_{\alpha}$ that might be hard to satisfy with finite-dimensional matrices. As mentioned in the main text, satisfying the original condition in Eq.~\eqref{eq:regcondition}, as imposed by the finiteness of the kinetic energy, should be sufficient for most practical applications. 

\end{document}